\documentclass[fleqn,usenatbib]{mnras}

\usepackage{newtxtext,newtxmath}

\usepackage[T1]{fontenc}

\DeclareRobustCommand{\VAN}[3]{#2}
\let\VANthebibliography\thebibliography
\def\thebibliography{\DeclareRobustCommand{\VAN}[3]{##3}\VANthebibliography}

\usepackage{graphicx}	
\usepackage{amsmath}	
\usepackage{float}
\usepackage{placeins}
\usepackage{multirow}
\usepackage{threeparttable}
\usepackage{natbib}
\usepackage{rotating}
\usepackage{tikz}
\usepackage{caption}
\usepackage{subcaption}

\newcommand*\mean[1]{\bar{#1}}

\title[Regular rotation and low turbulence at \textit{z} $\sim$ 4.5]{Regular rotation and low turbulence in a diverse sample of \textit{z} $\sim$ 4.5 galaxies observed with ALMA}

\author[Roman-Oliveira et al.]{
Fernanda Roman-Oliveira$^{1}$\thanks{E-mail: romanoliveira@astro.rug.nl},
Filippo Fraternali$^{1}$,
Francesca Rizzo$^{2,3}$
\\
$^{1}$Kapteyn Astronomical Institute, University of Groningen, Landleven 12, 9747 AD, Groningen, The Netherlands\\
$^{2}$Cosmic Dawn Center (DAWN), Jagtvej 128, DK2200, Copenhagen N, Denmark \\
$^{3}$Niels Bohr Institute, University of Copenhagen, Lyngbyvej 2, DK-2100 Copenhagen Ø, Denmark
}

\date{Accepted XXX. Received YYY; in original form ZZZ}

\pubyear{2022}

\begin{document}
\label{firstpage}
\pagerange{\pageref{firstpage}--\pageref{lastpage}}
\maketitle

\begin{abstract}
The discovery of galaxies with regularly rotating discs at redshifts $\geq$ 4 has been a puzzling challenge to galaxy formation models that tend to predict chaotic gas kinematics in the early Universe as a consequence of gas accretion, mergers and efficient feedback. In this work, we investigated the kinematics of five highly resolved galaxies at \textit{z} $\sim$ 4.5 observed with ALMA in the [CII] 158 $\mu$m emission line.
The sample is diverse: AzTEC1 (starburst galaxy), BRI1335-0417 (starburst and quasar host galaxy), J081740 (normal star-forming galaxy) and SGP38326 (two starburst galaxies in a group).
The five galaxies show velocity gradients, but four were found to be rotating discs while the remaining, AzTEC1, is likely a merger.
We studied the gas kinematics of the discs using \texttt{$^{\text{3D}}$BAROLO} and found that they rotate with maximum rotation velocities between 198 and 562 km s$^{-1}$ while the gas velocity dispersions, averaged across the discs, are between 49 and 75 km s$^{-1}$. The rotation curves are generally flat and the galaxies have ratios of ordered-to-random motion ($V/\sigma$) between 2.7 and 9.8.
We present \texttt{CANNUBI}, an algorithm for fitting the disc geometry of rotating discs in 3D emission-line observations prior to modelling the kinematics, with which we find indications that these discs may have thicknesses of the order of 1 kpc.
This study shows that early disc formation with a clear dominance of rotation with respect to turbulent motions is present across a variety of galaxy types.


\end{abstract}

\begin{keywords}
galaxies: evolution -- galaxies: high-redshift -- galaxies: kinematics and dynamics --  submillimetre: galaxies
\end{keywords}


\section{Introduction}

The Atacama Large Millimiter/submillimetre Array (ALMA) has been fundamental for the exploration of the high-\textit{z} Universe in the past decade \citep{hodge20}. Through the observation of [CII] 158 $\mu$m emission at \textit{z} $ > 4$, the study of cold gas kinematics has opened a new window in our understanding of galaxy formation in the early Universe \citep{carilli&walter13}. Recent observations show numerous kinematically-cold disc galaxies with ratios of rotation-to-random motion ($V/\sigma$) as high as 10, some of them with bulges or spiral arms when the Universe was no older than 1.5 billion years \citep{hodge12, sharda19, neeleman20, rizzo20, lelli21, fraternali21, rizzo21, tsukui21}.

Meanwhile, state-of-the-art cosmological simulations struggle to reproduce these observations, as gas discs with $V/\sigma \geq 3$ usually form at \textit{z} $\leq$ 2 \citep{simons17,pillepich19}. The current theoretical framework of galaxy evolution predicts that the early stages of galaxy formation are chaotic and dominated by violent gas accretion and frequent mergers \citep[e.g.][]{dekel09, somerville15}. Furthermore, the interstellar medium (ISM) of galaxies is also expected to be significantly affected by feedback processes from either supernovae or an active galactic nucleus (AGN) which heat and eject the gas from the galaxy, therefore increasing the turbulent motions \citep{biernacki18, nelson19, hayward17}. The effects of stellar feedback on the ISM should become especially important for high-\textit{z} starburst galaxies, that can reach star formation rates (SFR) of the order of a thousand solar masses per year, posing difficulties in maintaining stable gas discs over large periods of time \citep{nelson19}.

Recent efforts to bridge the tension between observations and theory have been carried out. \citet{kretschmer22} found an example of a simulated cold gas disc at \textit{z} $\sim$ 3.5 with V/$\sigma$ similar to the observed values. However, such a disc lasts for only 5 orbital periods and returns to a highly disturbed state after only 400 Myr. In contrast, \citet{tamfal22} showed with a cosmological, N-body hydrodynamical simulation that a present-day Milky Way-type galaxy can form a thin stellar disc as early as \textit{z} $=8$ that remains kinematically cold. Similarly, \citet{kohandel20} presented simulated galaxies at \textit{z} = 6 $-$ 8 with V/$\sigma \sim 7$ and velocity dispersions in the range of 23 $-$ 38 km s$^{-1}$. Finally, some studies suggested that the formation and survival of a disc may depend on the mass of the dark matter hosting the galaxy with a threshold of $2\times10^{11}$ M$_{\odot}$, however, there is a lack of observations of low mass galaxies at high-\textit{z} to validate this hypothesis \citep{dekel20, gurvich22}.

Furthermore, the cosmic evolution of gas velocity dispersion in galaxies is still largely unconstrained. Determining the velocity dispersion of discs across cosmic time and the presence of regular rotation is essential to link early discs with their evolved counterparts in the Local Universe.
The general expectation from cosmological simulations and \textit{z} $\gtrsim 3$ observations is that velocity dispersions increase up to $\sim$ 100 km s$^{-1}$ at \textit{z} $> 4$ \citep{uebler19, pillepich19}, considerably larger than current observations indicate \citep{sharda19, neeleman20, rizzo20, lelli21, fraternali21, rizzo21, tsukui21}.
A precise quantification of such dispersions is essential to identify the mechanisms that drive gas turbulence in different stages of galaxy evolution. While some studies find that gravitational instabilities are necessary \citep{krumholz18, uebler19}, some other studies find that the relatively low turbulence found in high-\textit{z} galaxies can be driven by stellar feedback with no need of additional sources \citep{rizzo21, fraternali21}.
Although, given that most of the highly-resolved dynamically cold disc galaxies observed at \textit{z} $>4$ are massive dusty star-forming galaxies, the broad picture of how the disc turbulence is regulated in typical main-sequence and/or low-mass galaxies remains uncertain.

The main challenges of analysing the kinematics of galaxies at \textit{z} $\geq 4$ are the relatively low angular resolution obtained in observations compared to the small angular size of the galaxies and the low signal-to-noise ratio (SNR) due to cosmological dimming.
A powerful way of circumventing these problems is to observe high-\textit{z} galaxies in the [CII] fine-structure transition at 158 $\mu$m, which is a major gas coolant that traces multiple phases of the ISM due to its low ionisation potential \citep{gullberg15, tarantino21, ramospadilla21}. [CII] is the brightest fine-structure line in star-forming galaxies and it is practically unaffected by the cosmic microwave background attenuation \citep{lagache18}.
Since \citet{capak15} showed that [CII] is abundant in early galaxies and more extended than the dust and CO emission \citep{gullberg18,tadaki19}, [CII] has been fundamental to study the ISM and dynamics of high-\textit{z} galaxies.
In galaxies at \textit{z} = $4 - 5$, the effective radii of the [CII] distributions typically vary in the range 0.3" to 1.1", which translates to about $2 - 7$ kpc \citep{gullberg18, fujimoto20, lefevre20}.
To be able to measure the shape of a rotation curve and a reliable value of the gas velocity dispersion, it is imperative to have at least a few independent resolution elements across the major axis of the disc. Thus, even for the most extended galaxies an angular resolution of at least 0.3" is necessary.

To elucidate some of the above questions, we have obtained accurate [CII] kinematic measurements of a sample of \textit{z} $\sim$ 4.5 galaxies with the 3D tilted-ring kinematic modelling tool \texttt{$^{\text{3D}}$BAROLO} \citep{diteodoro15}.
We used publicly available high-resolution [CII] data, some of which have been presented before, but, together they have not been systematically studied with a methodology that is non-parametric, corrected for beam smearing and in 3D.
This paper focuses on the presentation of the data and the kinematic modelling.
A following paper will undertake the dynamical modelling analysis, the local disc stability \citep{toomre64} and a study on the origin of the gas turbulence. This paper is structured as follows: in Section~\ref{sec:data}, we present the sample selection and data reduction. In Section~\ref{sec:methods}, we present the methodology for the 3D kinematic modelling and the disc geometry modelling. In Section~\ref{sec:results}, we show our main results: the geometry of the galaxies and the kinematic properties derived. In Section~\ref{sec:discussion} we interpret our findings within the context of previous works. In Section~\ref{sec:conclusions} we summarise our main findings and implications. Throughout the paper, we adopt a $\Lambda$CDM cosmology with Hubble constant $H_0$ = 67.7 km/s/Mpc, matter density $\Omega_{\text{m}}$ = 0.31 and vacuum energy $\Omega_{\lambda}$ = 0.69, from the 2018 Planck results \citep{planck20}.

\section{Data}\label{sec:data}

To compile our sample of highly spatially resolved galaxies, we searched the ALMA archive for public datasets with [CII] 158 $\mu$m line emission. Our selection criteria were the following: observations with angular resolution $<0.3$"; spectral resolution $<30$ km s$^{-1}$; redshift of the galaxy in the range of \textit{z} $ = 4-5$.
This is a particularly interesting redshift range as it lies in between the Epoch of Reionisation and cosmic noon, two pivotal points in galaxy evolution. Furthermore, this redshift range has the benefits of: having the redshifted [CII] 158 $\mu$m emission-line conveniently placed at an atmospheric window covered by ALMA; being beyond the cosmic peak of angular diameter distance, which allows for better spatial resolution than objects at cosmic noon; and having lower cosmological dimming than at higher \textit{z} \citep{carilli&walter13}.
We avoided selecting galaxies in massive proto-clusters as they may have a more evolved dynamics than typical \textit{z} $\sim 4.5$ galaxies and we avoided lensed sources as they require a different analysis as the one in \citet{rizzo18, rizzo20, rizzo21}.
To perform this search we used the tool \texttt{astroquery} \citep{ginsburg19} and cross-referenced it with the NED database to obtain an estimate of the redshift.
The above criteria resulted in four publicly available datasets.

\begin{table*}
\begin{center}
\caption{Summary of the properties of the sample. Columns from left to right: (1) ID of the source. (2) and (3) The coordinates represent the centre of the galaxies as used in the kinematic modelling in Section~\ref{sec:kin}, and correspond to the centres given in Table~\ref{tab:BBpar} (see Appendix~\ref{app:kinpar}) with typical errors of 1 pixel, which corresponds to 0.02 -- 0.03". (4) Redshift from the central velocity obtained with \texttt{$^{\text{3D}}$BAROLO} from the [CII] global line profile, the errors are dependent on the spectral resolution of the dataset, but are typically about $5 \times 10^{-4}$. (5) Physical scale at the given redshift. (6) Channel widths. (7) Angular resolution of the cleaned datacubes. (8) The root mean square (RMS) noise per velocity channel ($\sigma_{\mathrm{[CII]}}$). (9) [CII] line total intensity. (10) Total integration time of the datasets used.}\label{tab:data}
\begin{tabular}{lccccccccccc}
\hline \hline
\multirow{2}{*}{ID}    & \multirow{2}{*}{RA}        & \multirow{2}{*}{DEC}        & \multirow{2}{*}{\textit{z}}   & Physical Scale & Channel width    & [CII] Beam  & [CII] RMS & I$_{\textrm{[CII]}}$ & int time \\
 &  &  &  & (kpc/") & (km s$^{-1}$) &  (")  &  (mJy/beam)  & (Jy km s$^{-1}$) & (s) \\ \hline \vspace{1.2mm}
AzTEC 1 $^{\mathrm{a}}$      & 09:59:42.86 & +02:29:38.18 & 4.3419 & 6.86 & 30 & 0.15 $\times$ 0.13 & 0.41 & 25.5 $^{+1.0}_{-1.2}$ & 1935 \\ \vspace{1.2mm}
BRI1335-0417 $^{\mathrm{b}}$ & 13:38:03.42 & -04:32:34.98 & 4.4071 & 6.82 & 15 & 0.20 $\times$ 0.17 & 0.43 & 38.8 $^{+1.2}_{-0.9}$ & 3629  \\ \vspace{1.2mm}
J081740 $^{\mathrm{c}}$      & 08:17:40.87 & +13:51:38.23 & 4.2603 & 6.92 & 26 & 0.24 $\times$ 0.16 & 0.35 & 6.5 $^{+0.4}_{-0.6}$ & 5988  \\ \vspace{1.2mm}
SGP38326-1 $^{\mathrm{d}}$   & 00:03:07.23 & -33:02:50.61 & 4.4238 & 6.80 & 26 & 0.18 $\times$ 0.16* & 0.19 & 7.9 $^{+0.4}_{-0.5}$ &  16844  \\ \vspace{0.4mm}
SGP38326-2 $^{\mathrm{d}}$   & 00:03:07.12 & -33:02:51.15 & 4.4274 & 6.80 & 26 & 0.18 $\times$ 0.16* & 0.19 & 1.9 $^{+0.1}_{-0.1}$  & 16844  \\\hline
\end{tabular}
\begin{tablenotes}
  \item * Angular resolution achieved with \texttt{tclean}/\texttt{uvtaper} set to 0.1".
  \item ALMA project number: $^{\mathrm{a}}$ 2017.1.00127.S (PI: Iono); $^{\mathrm{b}}$ 2017.1.00394.S (PI: Gonz\'alez L\'opez); $^{\mathrm{c}}$ 2017.1.01052.S (PI: Neeleman); $^{\mathrm{d}}$ 2015.1.00330.S (PI: Riechers), 2018.1.00001.S (PI: Oteo).
\end{tablenotes}
\end{center}
\end{table*}

In Table~\ref{tab:data}, we present the properties of the [CII] line emission and datasets of our galaxy sample.
In Table~\ref{tab:masses}, we present the SFRs and molecular gas masses as previously reported in the literature for these objects. 
We estimate the gas masses based on high transition CO luminosities reported in the literature as referenced in Table~\ref{tab:masses}. For AzTEC 1, BRI1335-0417, SGP38326 (both galaxies), we use the J $= 4-3$, $2-1$ and $5-4$ transitions from \citet{yun15, jones16, oteo16}, respectively. We estimate the the L'$_{\textrm{CO(J=1-0)}}$ using the line ratios from \citet{carilli&walter13} and we assume an $\alpha_{\textrm{CO}}$ of 0.8 M$_{\odot} (\textrm{K km s}^{-1} \textrm{pc}^2)^{-1}$ to estimate the molecular gas masses. For J081740, a normal star-forming galaxy, we follow the assumptions from \citet{neeleman20} and report the gas mass obtained with a line ratio of 0.81 (between J $=2-1$ and J $=1-0$) and an $\alpha_{\textrm{CO}}$ of 3 M$_{\odot} (\textrm{K km s}^{-1} \textrm{pc}^2)^{-1}$, appropriate for a normal (non-starburst) high-\textit{z} star-forming galaxy.
As for the SFRs, we normalise them all to the initial mass function (IMF) of \citet{chabrier03} according to \citet{madau&dickinson14}.

\begin{table}
\begin{center}
\caption{Star formation rates and gas masses of the sample as previously reported in the literature. The SFR are normalised to a Chabrier IMF \citep{chabrier03}. In the case of BRI1335-0417 the references 2 and 3 are for the SFR and gas mass, respectively.}\label{tab:masses}
\begin{tabular}{lccc}
\hline \hline
ID & SFR (M$_{\odot}$ yr$^{-1}$) & M$_{\mathrm{H_2}}$ (M$_\odot$) & References \\ \hline
AzTEC 1      & $1240 \pm 230$ & $1.4 \pm 0.2 \times 10^{11}$ & 1 \\
BRI1335-0417 & $5100 \pm 1500$ & $1.0 \pm 0.1 \times 10^{11}$ & 2, 3 \\
J081740      & $166 \hspace{1mm} ^{+344}_{-103}$  & $8.8 \pm 2.6 \times 10^{10}$ & 4 \\ 
SGP38326-1   & $\sim1830$ & $\sim1.9 \times 10^{11}$ & 5 \\
SGP38326-2   & $\sim882$ & $\sim7.6 \times 10^{10}$ & 5 \\ \hline
\end{tabular}
\begin{tablenotes}
  \item 1. \citet{yun15}; 2. \citet{lu18}; 3. \citet{jones16}; \\
  4. \citet{neeleman20}; 5. \citet{oteo16}.
\end{tablenotes}
\end{center}
\end{table}

Here we summarise the main properties of each galaxy:

\begin{itemize}
    \item AzTEC 1 \\ 
        A sub-millimetre galaxy characterised by a clumpy dust continuum \citep{iono16}. It has a SFR of $\sim$1300 M$_{\odot}$ yr$^{-1}$ and an estimated stellar mass of $\sim10^{11}$M$_{\odot}$, derived from fitting the spectral energy distribution (SED) over the range of rest-frame UV to radio wavelengths \citep{yun15}. It was observed in multiple tracers with ALMA, such as CO(J=4-3), [NII] 205 $\mu$m and [OIII] 88 $\mu$m in addition to the [CII] 158 $\mu$m, analysed in this paper. The L$_{\text{[OIII]}}$/L$_{\text{[NII]}}$ suggests a chemically evolved system with a gas metallicity of 0.7-1.0 Z$_{\odot}$ \citep{tadaki19}. The high-resolution [CII] dataset of AzTEC1, used in this paper, was previously analysed by \citet{tadaki20} with GalPaK$^{\mathrm{3D}}$, a parametric kinematic modelling tool for 3D data \citep{bouche15}. They found a rotating disc with velocity dispersion of $\sim$ 70 km s$^{-1}$ and reported the presence of non-corotating gas components which they interpreted as an infall that is taking place perpendicularly to the disc. 
    \item BRI1335-0417 \\ 
        An infrared hyperluminous quasar host galaxy selected by the Automated Plate Measuring (APM) optical survey for high-\textit{z} quasars \citep{irwin91}. It was one of the first CO detections at \textit{z} $>$ 4, observed in the CO(J=5-4) line with IRAM \citep{guilloteau97}, later also observed in [CII] with the APEX telescope \citep{wagg10} and with ALMA at high-resolution. Recently, using the [CII] emission, \citet{tsukui21} found a structure suggestive of the presence of spiral arms together with the dynamical evidence of a massive bulge in the centre. CO(J=7-6) and [NII] observations showed the existence of a possible outflow \citep{lu18}. BRI1335-0417 is an extreme starburst with a SFR estimated to be around 5000 M$_{\odot}$ yr$^{-1}$. \citet{wagg14} found this estimate via SED fitting, which may, however, be biased by the presence of an AGN. Later, \citet{lu18} found a similar estimate with a method based on the luminosity of the CO(J=7-6) emission that is supposed to be independent of the presence of an AGN because it is insensitive to AGN heating \citep{lu17}. The high-resolution [CII] dataset of BRI1335-0417, used in this paper, was previously analysed by \citet{tsukui21} with the KinMS package, a parametric kinematic modelling tool for 3D data \citep{davis20, davis13}. They reported a velocity dispersion of $\sim$ 70 km s$^{-1}$ in the outer parts of the galaxy.
        \item J081740 \\ 
         A galaxy first identified as a damped Lyman alpha system in the spectrum of a background quasar \citep{neeleman17}. Because this selection method is not biased towards objects brighter in the submillimetre, it is the only non-starburst galaxy in this sample. With a SFR of only 177 M$_{\odot}$ yr$^{-1}$, which is likely representative of more typical star-forming galaxies at \textit{z} $\sim$ 4. The high-resolution [CII] dataset of J081740, used here, was previously analysed by \citet{neeleman20} with QubeFit, a parametric kinematic modelling tool for 3D data \citep{neeleman20qubefit}. They showed the presence of a rotating disc with a velocity dispersion of $\sim$ 80 km s$^{-1}$.
    \item SGP38326 \\ 
          A pair of starbursts with a combined SFR of around 4500 M$_{\odot}$ yr$^{-1}$, here we refer to the more massive galaxy as SGP38326-1 and the companion as SGP38326-2. Detection in $^{12}$CO(J=5-4) showed an elongated emission connecting the two galaxies which could be interpreted as a molecular gas bridge among them suggesting an ongoing interaction \citep{oteo16}. However, dust was not detected along this region and there appear to be no signs of stripped material in the [CII] emission \citep{oteo16}. This system is considered one of the best examples of high-redshift starburst galaxies that can be the progenitor of massive elliptical galaxies at lower redshifts \citep{valentino20, toft14}. A lower-resolution [CII] observation was analysed by \citet{oteo16}, where they fit the observed velocity field with a simple dynamical model.
\end{itemize}

Since we selected galaxies based solely on the high quality of the data available we are analysing the kinematic properties of a diverse set of systems: starburst galaxies isolated and in a group, a quasar host galaxy and a normal star-forming galaxy.

\subsection{Data Reduction and Imaging}\label{sec:datared}

After retrieving the raw data from the ALMA archive, we carried out the data reduction using the Common Astronomy Software Applications (CASA) \citep{mcmullin07}. This was done in two steps: 1) calibration of the visibilities and 2) imaging of the line emission datacubes and dust continuum maps.

In the first step, we calibrated the visibility data to obtain measurement sets in which we could begin the imaging procedure. For that, we used the original calibration scripts with the same CASA version in which the data had been taken. This process was straightforward for the sources AzTEC1, BRI1335-0417 and J081740. However, for SGP38326 we combined two separate high-resolution observations to achieve a higher SNR (Project IDs: 2018.1.00001.S and 2015.1.00330.S). We used the task \texttt{split} to average both datasets to a similar channel width and, subsequently, we used the task \texttt{concat} to concatenate the datasets to a unique measurement set used for the imaging. We only used the frequency range common to both datasets.

In the second step, the imaging procedure was done using the CASA version 5.6.1-8 and we started by first producing a dirty cube in which we identified the velocity channels containing line emission. To obtain a continuum map, we first produced a dirty continuum image of the line-free channels using the task \texttt{tclean} in the Multi Frequency Synthesis (MFS) mode to estimate the root mean square (RMS) of the noise in the continuum map. Subsequently, we produced the final clean continuum map using \texttt{tclean} in MFS mode and cleaning down to 3 times the RMS noise of the dirty image for the sources AzTEC1, BRI1335-0417 and J081740. In the case of SGP38326 group, we chose to clean it further towards 1.5 times the RMS noise of the dirty image so we could probe fainter substructures that may be present in the data.

To obtain a line-emission datacube, we first created a continuum-free measurement set using the task \texttt{uvcontsub} with a linear fit for the sources AzTEC1, BRI1335-0417 and J081740. Similarly to the continuum map, we first created a dirty line-emission cube using \texttt{tclean} in the 'cube' mode to find the RMS noise level per velocity channel and then produced a final line-emission cube cleaned down to 3 times the RMS noise.

For SGP38326, we followed a slightly different procedure: since it is a complex system with many continuum sources, we avoided using \texttt{uvcontsub} to subtract the continuum. We created a dirty cube of the continuum and line emission of the system with \texttt{tclean} to estimate the RMS noise per velocity channel, we then created a clean cube of the continuum and line emission cleaning it down to 1.5 times the RMS noise of the dirty cube. To extract the final line-emission cube and continuum map we used the task \texttt{imcontsub} that removes the continuum from the cube in the image space rather than in the uv space. For this specific dataset, we also used \texttt{uvtapering} in \texttt{tclean} to taper the spatial resolution to 0.1" to maintain a suitable SNR while slightly compromising the maximum possible angular resolution of the dataset.

During the imaging process, we produced emission-line cubes with Briggs weighting \citep{briggs95} with a robust parameter of 2 for the sources AzTEC1, J081740 and SGP38326 which is equivalent to a natural weighting. For BRI1335-0417, since the SNR is high enough, we chose a robust parameter of 0.5 to achieve a slightly better spatial resolution. As for the dust continuum maps, we also used a robust parameter of 2 for the sources J081740 and SGP383260, although, AzTEC1 and BRI1335-0417 are bright enough sources that we could use a robust parameter of 0.5 to achieve a better spatial resolution. Finally, we chose cell sizes that are within 5 to 8 times smaller than the beam size: 0.02"/px for AzTEC1 and SGP38326; and 0.03"/px for BRI1335-0417 and J081740.

Our choices for the weighting schemes and channel widths were motivated to obtain an average SNR of at least 4 per velocity channel in at least 10 channels of the data cubes. We define the SNR of the line emission as the ratio between the average of the masked [CII] line intensity per pixel in each velocity channel and the RMS noise. This is the masking used in the kinematic modelling that we describe in more detail in Sections~\ref{sec:barolo} and ~\ref{sec:presentation}.

\section{Methodology}\label{sec:methods}

\subsection{Modelling the disc kinematics with \texttt{$^{\text{3D}}$BAROLO}}\label{sec:barolo}
\texttt{$^{\text{3D}}$BAROLO} is an algorithm that estimates the best-fit parameters of a rotating disc that reproduce the emission in the observed emission-line datacubes \citep{diteodoro15}. \texttt{$^{\text{3D}}$BAROLO} minimises the residuals between the observations and the mock datacubes produced with 3D tilted-ring models and convolved to the same resolution of the observations. Therefore, it accounts for the effect of beam smearing \citep{bosma78, begeman87}, which is very important for retrieving robust measurements of velocity dispersions and rotation velocities at the relatively low angular resolution of the present data.

For the fits, we assume that the emission is tracing circular motions and that there are no large-scale radial or vertical motions, e.g. inflows/outflows, although we discuss this further in Section~\ref{sec:kin} where we also note that some galaxies present some distortions that could be interpreted as non-circular motions.

Before performing the 3D modelling, we mask the data by performing a source detection in 3D.
This masking procedure is handled in \texttt{$^{\text{3D}}$BAROLO} by the task \texttt{SEARCH} with two parameters: \texttt{SNRCUT} a primary cut the emission at a chosen SNR level; and \texttt{GROWTHCUT} a secondary SNR boundary that limits the growth of the mask.
The input disc geometry parameters are explained in the Section~\ref{subsec:CANNUBI}.

Moreover, we strive to not oversample the rotation curves thus we keep the number of modelled tilted rings as close as possible to the real number of resolution elements in the data with a separation close to the angular resolution. As a consequence all our measured rotation velocities and velocity dispersions in a galaxy can be considered independent. The \texttt{$^{\text{3D}}$BAROLO} parameters used for the kinematic fits are given in Table~\ref{tab:BBpar}, Appendix~\ref{app:kinpar}. In Section~\ref{sec:results}, we present the best-fit kinematic models for the sample.

\subsection{Estimating the disc geometry}\label{subsec:CANNUBI}

Modelling the kinematics of a rotating disc with \texttt{$^{\text{3D}}$BAROLO} requires a set of geometric parameters for input. To reduce the number of free parameters in the kinematic fitting we derive the geometric parameters beforehand using \texttt{CANNUBI}\footnote{Available at: \url{https://www.filippofraternali.com/cannubi}}, a Markov Chain Monte Carlo (MCMC) python routine for estimating the disc geometry of galaxies.

\texttt{CANNUBI} uses \texttt{emcee} \citep{mackey13} to minimise the data against resolution-matched 3D tilted-ring models of rotating discs generated with \texttt{$^{\text{3D}}$BAROLO}.
\texttt{CANNUBI} estimates the coordinates of the centre of the disc, its radial extent, its disc thickness (Z0) and the position angle (PA) and inclination angle.
The minimisation is done on the total flux map of the galaxy or in 3D using the full data cube. The latter can be useful to overcome the degeneracy between inclination and thickness as the data cube holds more information on the geometry, however, it is much more computationally demanding as the code has to perform a full kinematic fit at every iteration. In this work we present tests done on the total flux map of mock galaxies. For more information on \texttt{CANNUBI} we refer the reader to the description linked in the Data Availability section.

Of all the geometric parameters measured with \texttt{CANNUBI}, the accurate determination of the disc inclination is key to a successful kinematic model, since it affects the correction from line-of-sight to rotation velocities. Inclinations of gaseous galaxy discs are often derived assuming a razor-thin disc and therefore that the shape of the emission on the sky is fully due to the orientation of an intrinsically round disc. In this case, the axis ratio is a direct measurement of the cosine of the inclination angle.
Another common assumption of disc thickness is to consider an inclination angle of $\cos^2 i = ((b/a)^2-q_0^2)/(1-q_0^2)$, where \textit{a} and \textit{b} are the semi-major and semi-minor axes, respectively. The disc thickness is taken into account in the parameter $q_0$ with a typical assumed value of $q_0 = 0.2$ for \textit{z} $\approx 2$ galaxies \citep{forster18, swinbank17, weijmans14}.

To obtain the fiducial kinematic models we discuss in Section~\ref{sec:results}, we used \texttt{CANNUBI} to estimate the disc inclination of the galaxies by assuming razor-thin discs.
However, due to high velocity dispersions, it is reasonable to expect that high-\textit{z} galaxies have non-negligible thickness, which can lead us to underestimate the disc inclinations \citep{iorio17}.
\texttt{CANNUBI} can potentially be used for measuring the thickness as shown by our tests on mock galaxies in Section~\ref{sec:CANNUBI}.
However, mock galaxies have idealised rotating discs that behave much more predictably than real galaxies.
Given the quality of the present data we are not fully confident that we can determine the disc thickness in our galaxies reliably.
Thus, we prefer to discuss our estimates in terms of uncertainty in our kinematic models in Section~\ref{sec:discthick}. It is important to note that if the thickness of the discs is smaller than the resolution of the data ($\sim$ 1 kpc), assuming zero thickness or a few hundred parsec would make no significant difference in our final geometrical and kinematic parameters.

\subsubsection{Creating a sample of mock galaxies}
To verify that \texttt{CANNUBI} can return reliable estimations of the inclinations and, to some extent, also the disc thickness, we performed a test run with mock data. We use a sample of 48 mock galaxies produced with the task \texttt{GALMOD} within \texttt{$^{\text{3D}}$BAROLO} that were created with a combination of several properties in the range of our real data.\texttt{GALMOD} creates mock models by populating a 3D space with gas clouds assuming circular orbits within a tilted ring model. We use a radial exponential surface density profile for the emitting gas and a Gaussian vertical density profile. We assume an empirical function for the rotation curve that is described by the following expression:

\begin{equation}
    V_{\mathrm{rot}}(R) = V_{\mathrm{t}} \hspace{2mm} \dfrac{\left(1 + \dfrac{R_{\mathrm{t}}}{R}\right)^{\beta}}{\left[1 + \left(\dfrac{R_{\mathrm{t}}}{R}\right)^{\xi}\right]^{\frac{1}{\xi}}}.
\end{equation}

This function is described in \citet[see Equation 20]{rizzo18} and in \citet{courteau97}. We chose the following numerical values for the parameters: $V_t=250$ km s$^{-1}$ for the velocity scale; $R_t=1$ kpc for the turnover radius between the central rising part and the outer part of the rotation curve; $\beta=0.4$ that is related to the behaviour at large radii and $\xi=2.5$ that specifies the sharpness of the turnover at small radii. These return a rising rotation curve in the centre and an external rotation curve that flattens with rotation velocity of 250 km s$^{-1}$. We assume a constant velocity dispersion of 50 km s$^{-1}$. We convolved the mock data with a synthetic beam to achieve 3 and 5 resolution elements per galaxy side with a galaxy radius of 3 kpc.

Moreover, we added random noise, convolved to the same resolution as the model, to achieve an average SNR per velocity channel of 3 and 5. The definition we use for SNR can be found in Section~\ref{sec:datared}.
We choose all these values to obtain mocks similar to the ALMA data we are analysing, i.e. with SNR between 4 and 5.5 and 3 to 5 independent resolution elements.
To make sure that we can properly estimate the inclination of the data we use a large range of inclinations: from 30 to 70 degrees. For the disc thickness we also probe a large range of possible values with a conservative choice for the largest possible thickness: from 0 to 1.6 kpc (constant with radius), in which the latter value would mean a disc as thick as half of its radial extent.
A summary of the properties used to create the sample of mock galaxies with a combination of all of these parameters can be found in Table~\ref{tab:mock}.



\begin{table}
\begin{center}
\caption{Properties of the mock galaxies used to test \texttt{CANNUBI}.}\label{tab:mock}
\begin{tabular}{lll}
\hline \hline
Parameter & Value & Unit \\
\hline 
N Sample                                & 48                        &  \\
Resolution elements per galaxy side     & 3, 5                      &  \\
SNR                                     & 3, 5                      &  \\
Galaxy radius                           & 3                         & kpc           \\
Rotation velocity                       & 250                       & km s${^{-1}}$ \\
Velocity dispersion                     & 50                        & km s${^{-1}}$ \\
Inclination                             & 30, 50, 70                & degrees \\
Disc thickness                          & 0, 0.2, 0.8, 1.6          & kpc \\ \hline
\end{tabular}
\end{center}
\end{table}

With \texttt{CANNUBI} we run an MCMC exploration on the total flux map of the galactic centre, radial extension, inclination and thickness with 20 walkers until convergence on all mock galaxies, which is enough to probe the parameter space of these mock discs. The convergence is estimated using the autocorrelation times of the parameters as explained in \citet{goodman10}. We analyse these results in the following Section comparing them to the input values. 

\subsubsection{Results on the mock data}\label{sec:CANNUBI}
\begin{figure}
    \centering
    \includegraphics[width=0.45\textwidth]{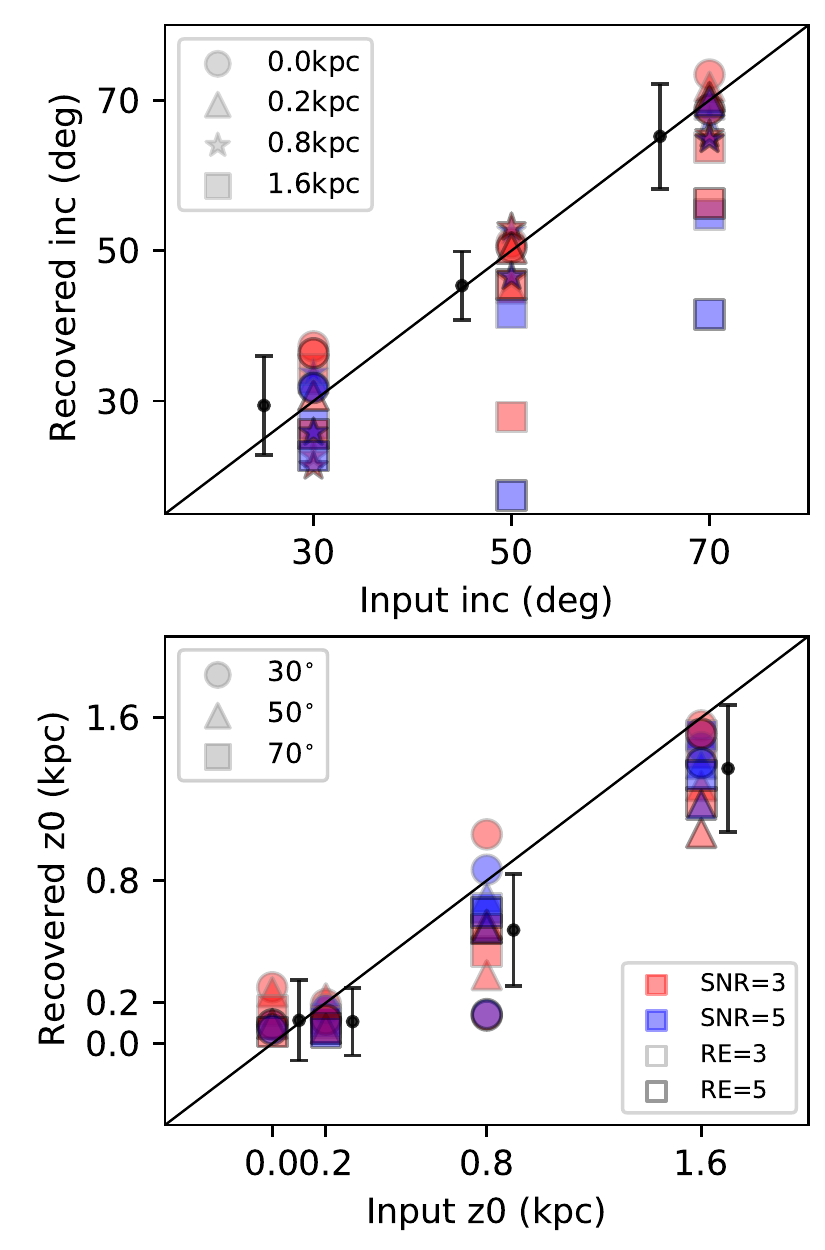}
    \caption{Recovery of the input inclination and disc thickness of the mock galaxies using the script \texttt{CANNUBI}. Top panel: recovered inclination vs. input inclination. All the \texttt{CANNUBI} runs are shown with markers according to the legend, representing different disc thicknesses: 0 kpc (circle), 0.2 kpc (triangle), 0.8 kpc (star) and 1.6 kpc (square).
    Bottom panel: recovered disc thickness vs. input disc thickness. All the \texttt{CANNUBI} runs are shown with markers representing different disc inclinations: 30$^{\circ}$ (circle), 50$^{\circ}$ and 70$^{\circ}$ (square). In both panels, we distinguish different SNR by color (red for SNR=3 and blue for SNR=5) and different amount of resolution elements per galaxy side by a marker edge (light gray for RE=3 and dark gray for RE=5). We show the typical errors as errorbars, these are the average standard deviations of the posteriors in each inclination group (top panel) and each disc thickness group (bottom panel). The black lines represent the identity line indicating perfect recovery.}
    \label{fig:alb_inc_mock}
\end{figure}

To validate the performance of \texttt{CANNUBI}, we show in Figure~\ref{fig:alb_inc_mock} the comparison between the input and recovered inclinations and disc thickness of the mock galaxies. We show the results over the three values of input inclinations and over the four values of input disc thickness and we show the average errors of each group of inclination and disc thickness as an errorbar.
The galactic centres of the mock galaxies are recovered on average within 1 pixel from the true centre and the disc extent is recovered on average within 0.1 kpc of the true extent. Here we concentrate on the recovered values of the inclination angle and the disc thickness.
We find that, on average, \texttt{CANNUBI} recovers the correct inclination within the errors for all groups of inclination: 29$^\circ \pm 7$, 45$^\circ \pm 5$ and 65$^\circ \pm 7$ for input inclinations of 30$^\circ$, 50$^\circ$ and 70$^\circ$, respectively.
The performance is slightly affected by the thickness of the discs, in which thicker input discs tend to have higher errors of up to 12$^\circ$, while the average errors for razor-thin discs are around 3$^\circ$.
The errors could change slightly depending on other factors that we are not exploring in this simple test, such as different velocity curves and surface brightness profiles.

Furthermore, in Figure~\ref{fig:alb_inc_mock} we can identify two main situations to be careful of: low-inclination discs and/or very thick discs. In the case of low-inclination discs, the disc thickness estimation can be overestimated as we reach the boundaries of the parameter space, although this is a marginal effect of about 0.1 kpc and it does not affect much the inclination estimates. In the case of very thick discs, the disc thickness is reasonably recovered at the cost of considerably underestimating the inclination, particularly for highly inclined discs (as seen in the outliers in the top panel of Figure~\ref{fig:alb_inc_mock}).
This is due to the fact that the shape of the total intensity map is significantly affected by the disc thickness. In the case of very thick discs, we did a preliminary investigation with \texttt{CANNUBI} in the full cube mode: in this setting we are minimising the disc geometry models over all the velocity channels instead of the total intensity map. Such a test showed promising results for the thickest discs, however, since it is very computationally intense we could not run it for all the mocks.

We would also like to note that, when the input is a zero thickness disc, we recover a thickness value that is slightly higher than zero. This is due to the fact that here we give the mean of the posterior distributions and that we do not probe regions of negative thickness. As a result, the mean and best-fit value do not fully match, although the distributions still peak at the true value.

Other than that, the disc inclination is, in general, well recovered and, therefore, we can conclude that \texttt{CANNUBI} satisfactorily constrains the inclination of mock galaxies within 30$^{\circ}$ to 70$^{\circ}$ and with disc thickness up to a quarter of the extent of the galaxies.
In the particular case of zero thickness, which we use for our galaxy sample (Section~\ref{sec:disc}), the experiments with the mock galaxies show an excellent recovery of the inclinations with errors of about 3$^{\circ}$.


\section{Results}\label{sec:results}

\subsection{Presentation of the data}\label{sec:presentation}

\begin{figure*}
    \centering
    \includegraphics[width=\textwidth]{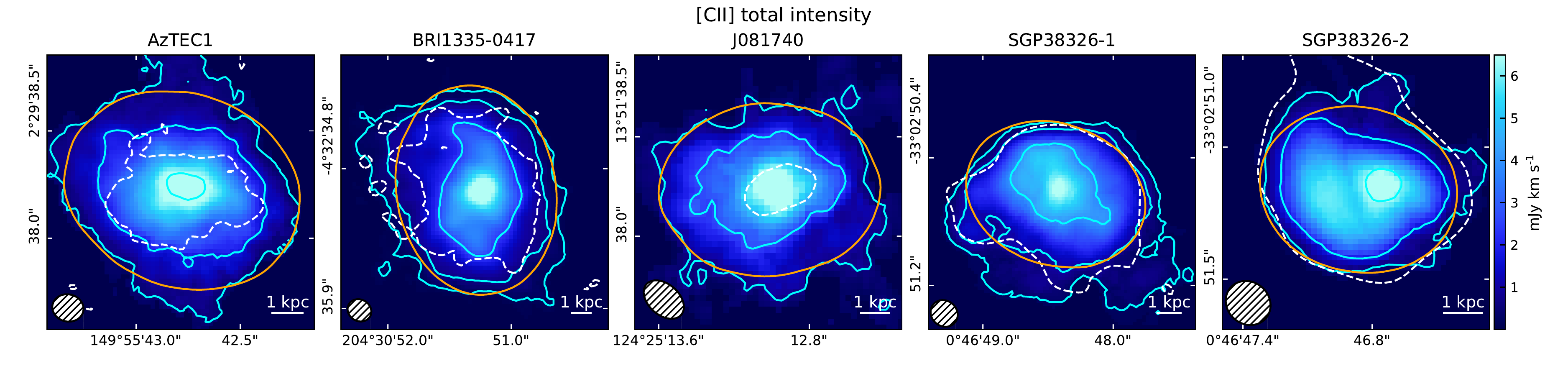}
    \caption{Intensity maps of the galaxies. The [CII] emission follows the colourmap and we show cyan contours in levels of emission that increase in powers of 4. The contours start from a pseudo 4$\sigma$-contour which corresponds to a flux of 1.7, 1.1, 1.4, 0.4 and 0.4 mJy km s$^{-1}$ for AzTEC1, BRI1335-0417, J081740, SGP38326-1 and SGP38326-2, respectively (see Appendix~\ref{app:ps} for more details). The dust emission is shown in white dashed contours at 3 times the continuum RMS noise. The beam and physical scale are shown in the bottom left and right of the panels, respectively. The orange contour shows our best-fit disc model obtained with \texttt{CANNUBI}, this is shown at the same level of emission as the pseudo 4$\sigma$-contour of the data.}
    \label{fig:data}
\end{figure*}

In Figure~\ref{fig:data}, we present the dust continuum and [CII] total intensity maps of all the galaxies in our sample. The [CII] total intensity maps were produced using \texttt{$^{\text{3D}}$BAROLO}, by collapsing the velocity range of the emission within the \texttt{SEARCH}-masked cube. We used the \texttt{SNRCUT} and \texttt{GROWTHCUT} parameters given in Table~\ref{app:kinpar}. Since the noise level of the mask-integrated emission map is different than the channel noise level and no longer uniform, we calculated a pseudo 4$\sigma$-contour ($\sigma_{\mathrm{pc}}$) as explained in Appendix~\ref{app:ps}. This is done for visualisation purposes and to ensure that the emission within this mask is real. The flux levels corresponding to the pseudo 4$\sigma$-contours are 1.7, 1.1, 1.4, 0.4 and 0.4 mJy km s$^{-1}$ for AzTEC1, BRI1335-0417, J081740, SGP38326-1 and SGP38326-2, respectively. They are shown as the outermost cyan contours in Figure~\ref{fig:data}, while the inner contours follow levels of flux above the pseudo 4$\sigma$-contour in powers of 4: $\sigma_{\mathrm{pc}}$, 4$\sigma_{\mathrm{pc}}$ and 16$\sigma_{\mathrm{pc}}$. We note that in the specific case of AzTEC1, there is some extended emission to the north and south with an unclear origin (e.g, cleaning artefacts, outflows).
Additionally, we also display the data cube channel maps in Appendix~\ref{app:cm}.

\begin{figure}
    \centering
    \includegraphics[width=0.45\textwidth]{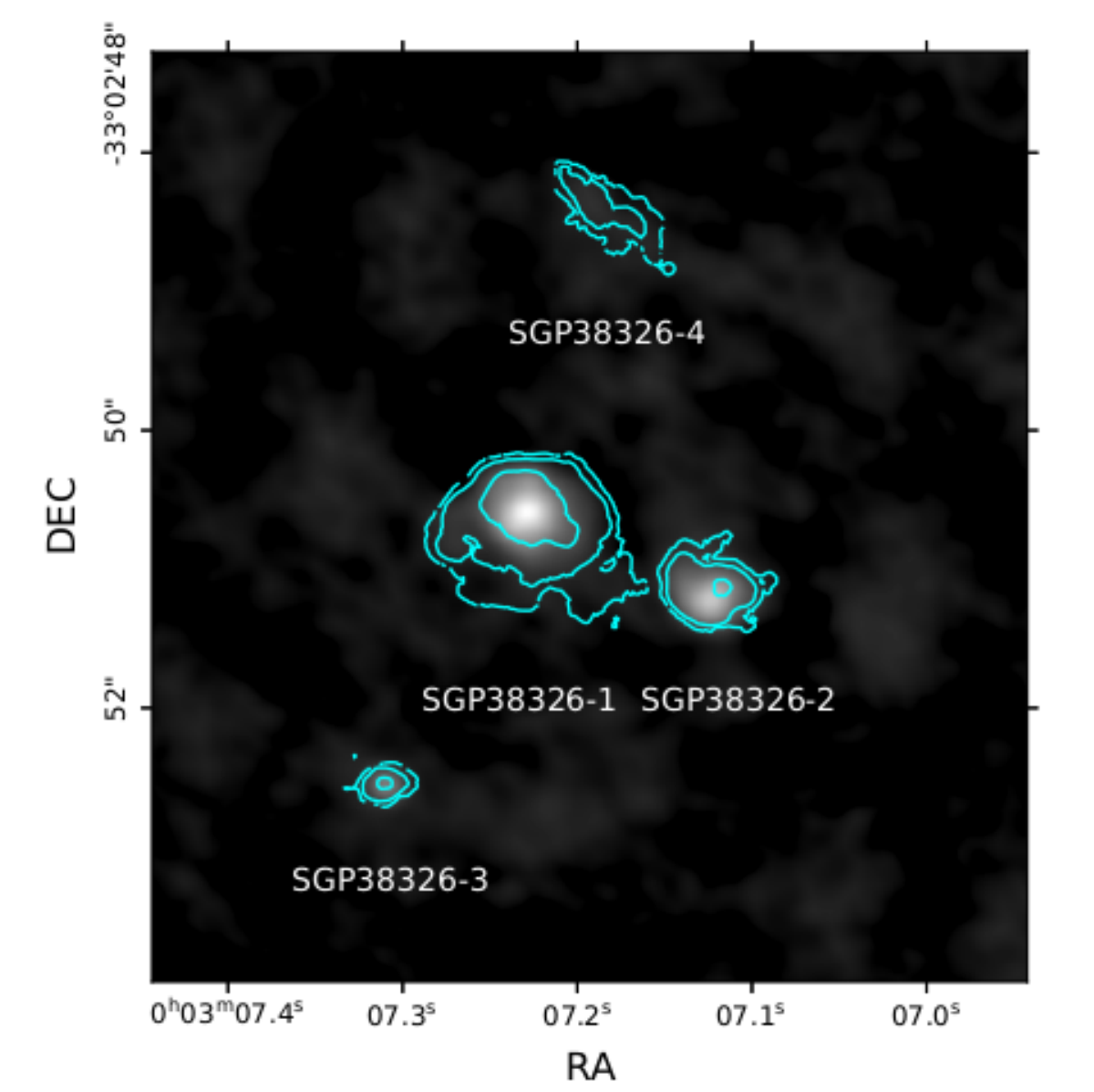}
    \caption{The SGP38326 group. Dust continuum is shown in gray scale with the masked [CII] emission shown in cyan contours with the same pseudo 4$\sigma$-contours as Figure~\ref{fig:data}. The RMS noise in the dust background is 0.06 mJy/beam. The white circles are for visual assistance only to show where the galaxies are.}\label{fig:sgp}
\end{figure}

In the system SGP38326, we are only analysing the two main galaxies that are well resolved. However, we also report the discovery of [CII] emission in two other companions in the system: SGP38326-3 and SGP38326-4, they are shown in Figure~\ref{fig:sgp}. The dust emission of SGP38326-3 was previously reported in \citet{oteo16}, however, the [CII] emission was not detected. We estimated the [CII] flux of SGP38326-3 and SGP38326-4 to be 0.35 and 0.16 Jy km s$^{-1}$, respectively, and we did not detect dust continuum in SGP38326-4 down to 0.06 mJy/beam, which is the value of the average noise in the image.
Both SGP38326-3 and SGP38326-4 are at an angular distance of about $\sim$2.0" or about 14 kpc in projection from the main galaxy SGP38326-1. SGP38326-3 and SGP38326-4 are at $\sim$600 km s$^{-1}$ and $\sim$800 km s$^{-1}$ spectral distance from the systemic redshift of SGP38326-1, respectively, as measured from the central velocity of both systems.
In spite of these [CII] emitters being initially discovered by visual inspection, we have also used the 3D source finder in \texttt{$^{\text{3D}}$BAROLO} and confirmed their presence. We recover the four objects using \texttt{SNRCUT} of 3.5 and a \texttt{GROWTHCUT} of 3.

\subsection{Kinematic modelling setup}
In this Section, we describe the setup we used in \texttt{$^{\text{3D}}$BAROLO} for the kinematic modelling of the sample alongside our assumptions made regarding the initial parameters for the geometry and kinematics.

\subsubsection{Disc geometry}\label{sec:disc}
Following the mock data tests we conducted in Section~\ref{sec:CANNUBI}, we used \texttt{CANNUBI} to estimate the disc geometry of the galaxies in our sample considering a razor thin disc for our fiducial kinematic models. We report the retrieved radial extent, inclination and morphological PA in Table~\ref{tab:geometricdata} and the corner plots can be found in Appendix~\ref{app:cannubicorner}. The possible effects of non-zero thickness in the kinematics are discussed in Section~\ref{sec:discthick}.

In Table~\ref{tab:geometricdata}, we make a distinction between morphological and kinematic PAs: the morphological PA was estimated from the [CII] total flux map with \texttt{CANNUBI}, while the kinematic PA was obtained with \texttt{$^{\text{3D}}$BAROLO}. \texttt{$^{\text{3D}}$BAROLO} initially obtains the first guess of the PA by finding the line that goes through the galactic centre and maximises the velocity gradient across the velocity field. Then to derive the final kinematic PA, we ran \texttt{$^{\text{3D}}$BAROLO} with a second fitting stage with a parameter regularisation keeping all other parameters fixed, except \texttt{VROT}, \texttt{VDISP} and \texttt{PA}. For the PA regularisation we chose a 0th order polynomial interpolation.

In an ideal scenario of a purely rotating disc, both PAs should coincide within the uncertainties. A large misalignment between morphological and kinematic position angles could be evidence that the velocity gradient is not due to a rotating disc but instead indicates the presence of interacting systems or outflow motions \citep{krajnovic06}.
For the galaxies BRI1335-0417, J081740, SGP38326-1 and SGP38326-2 the misalignments are compatible within about 2$\sigma$. The misalignments for these galaxies are smaller than 15$^{\circ}$, except for SGP38326-2 that has a misalignment of 36$^{\circ}$ although it has larger errors associated with it. For AzTEC1 the two values differ considerably outside the uncertainties, by 38$^{\circ}$. This could be due to the presence of a bright region offset by $\sim$ 200 km s$^{-1}$ from the central velocity that shifts the orientation of the total flux map in a different direction in the outer region. This region can be seen in the residual channel maps (see second panel, third row of Figure~\ref{fig:cmA}).
We investigated this further in Section~\ref{sec:pvsplit}, where we found other indications that AzTEC1 is likely an interacting system.

\begin{table}
\begin{center}
\caption{Geometric parameters of the sample estimated with \texttt{CANNUBI} under the assumption of a thin disc and minimising over the total flux map. PA$_{\textrm{morph}}$ is the morphological position angle from the total map and PA$_{\textrm{kin}}$ is the kinematic position angle from the velocity gradient, the latter was estimated using \texttt{$^{\text{3D}}$BAROLO}.}\label{tab:geometricdata}
\begin{tabular}{lcccc}
\hline \hline
ID & Radial extent & PA$_{\textrm{morph}}$ & PA$_{\textrm{kin}}$ & Inclination   \\ 
   & (kpc)         & (deg)                 & (deg)               & (deg) \\ \hline  \vspace{1mm}
AzTEC1       & $4.4 \hspace{0.5mm} \pm \hspace{0.5mm} 0.1$ & $251 \hspace{1mm} ^{+8}_{-8}$   & $289 \hspace{1mm} ^{+2}_{-2}$ & $39 \hspace{1mm} ^{+5}_{-6}$      \\ \vspace{1mm}
BRI1335-0417 & $5.8 \hspace{0.5mm} \pm \hspace{0.5mm} 0.1$ & $8   \hspace{1mm} ^{+7}_{-7}$   & $10  \hspace{1mm} ^{+9}_{-9}$ & $42 \hspace{1mm} ^{+3}_{-4}$      \\ \vspace{1mm}
J081740      & $4.4 \hspace{0.5mm} \pm \hspace{0.5mm} 0.1$ & $86  \hspace{1mm} ^{+6}_{-6}$   & $99  \hspace{1mm} ^{+1}_{-1}$ & $43 \hspace{1mm} ^{+3}_{-5}$      \\ \vspace{1mm}
SGP38326-1   & $4.1 \hspace{0.5mm} \pm \hspace{0.5mm} 0.1$ & $71  \hspace{1mm} ^{+9}_{-9}$   & *$86  \hspace{1mm} ^{+5}_{-5}$ & $42 \hspace{1mm} ^{+5}_{-6}$       \\ \vspace{1mm}
SGP38326-2   & $2.5 \hspace{0.5mm} \pm \hspace{0.5mm} 0.1$ & $83  \hspace{1mm} ^{+18}_{-18}$ & $119 \hspace{1mm} ^{+8}_{-8}$ & $41 \hspace{1mm} ^{+10}_{-14}$  \\ \hline
\end{tabular}
\begin{tablenotes}
  \item * We note that, for SGP38326-1, we let the PA free in the kinematic modelling and it ranges from 88$^{\circ}$ in the innermost ring to 71$^{\circ}$ in the outermost (as explained in Section~\ref{sec:SGP1}).
\end{tablenotes}
\end{center}
\end{table}

\subsubsection{Input parameters}

In the kinematic modelling setup, we used an azimuthal normalisation of the surface density and kept the rotation velocity and velocity dispersion parameters free and all other parameters fixed with the exception of the PA for SGP388326-1, as explained in Section~\ref{sec:SGP1}. For the masking we used the \texttt{SEARCH} task to obtain a tight mask around the emission without any smoothing, the parameters \texttt{SNRCUT} and \texttt{GROWTHCUT} used are given in Table~\ref{tab:BBpar}. We used the systemic redshift determined by \texttt{$^{\text{3D}}$BAROLO} which is estimated as the central frequency in the global line profile.

With respect to the geometric parameters, we used the galactic centre obtained with \texttt{CANNUBI} which we slightly adjusted in some cases by visually inspecting the velocity field map and the position-velocity (p-v) diagrams (by less than 2 pixels, i.e. significantly within the resolution of the data). The galactic centre coordinates obtained are reported in Table~\ref{tab:data}.
Similarly, we slightly adjusted the values of the radial separation between the rings to better correspond to the emission seen in the p-v diagrams, the parameters for the number of rings (\texttt{NRADII}) and radial separation (\texttt{RADSEP}) are given in Table~\ref{tab:BBpar}.
The disc PA is fixed to the kinematic PA to ensure that we cross the region of the highest line-of-sight velocities in the disc, see Table~\ref{tab:geometricdata}.
We set the disc thickness to an infinitely thin disc and fix the inclination to the one estimated by \texttt{CANNUBI}, see Table~\ref{tab:geometricdata}.

During the fit, we kept the PA fixed across the disc, except for SGP38326-1 as explained in Section~\ref{sec:SGP1}. For all galaxies, we fitted the kinematics using both sides of the rotation curve, except for BRI1335-0417 as explained in Section~\ref{sec:bri}.

\subsection{Kinematic models}\label{sec:kin}

In this Section, we show and describe the best-fit kinematic models for each galaxy.

In Figure~\ref{fig:kinmod}, we show the velocity fields of the data and the velocity fields extracted from the 3D tilted-ring model in the first and second rows, respectively. These maps were masked with the same pseudo 4$\sigma$-contour from Figure~\ref{fig:data}.
The rotating disc best-fit models were obtained with \texttt{$^{\text{3D}}$BAROLO} using the parameters displayed in Table~\ref{tab:BBpar}.
In the second row, we show the outer contour of the axis-symmetric model obtained with \texttt{CANNUBI} (in orange), as described in Section~\ref{sec:disc}, alongside the outer contour of the model map produced by \texttt{$^{\text{3D}}$BAROLO} (in black). The geometry of the two models differ as we favoured the kinematic position angles over the morphological ones for the kinematic fitting. Both are usually very similar for the galaxies, with the notable exception of AzTEC1, as discussed in Section~\ref{sec:disc}.
Each of the modelled rings are represented by the white circles along the major axis of the velocity fields.
In the third and fourth rows, we present the p-v diagrams along the major and minor axes, respectively. We show the data emission in gray scale and blue contours following levels of 2$\sigma_{\mathrm{[CII]}}$, 4$\sigma_{\mathrm{[CII]}}$, 8$\sigma_{\mathrm{[CII]}}$ and 16$\sigma_{\mathrm{[CII]}}$. The negative emission is shown as gray contours at $-2\sigma_{\mathrm{[CII]}}$. We show the model contours in red following the same levels.

Now, we detail what can be seen in these maps for each individual galaxy:


\begin{figure*}
    \centering
    \includegraphics[width=\textwidth]{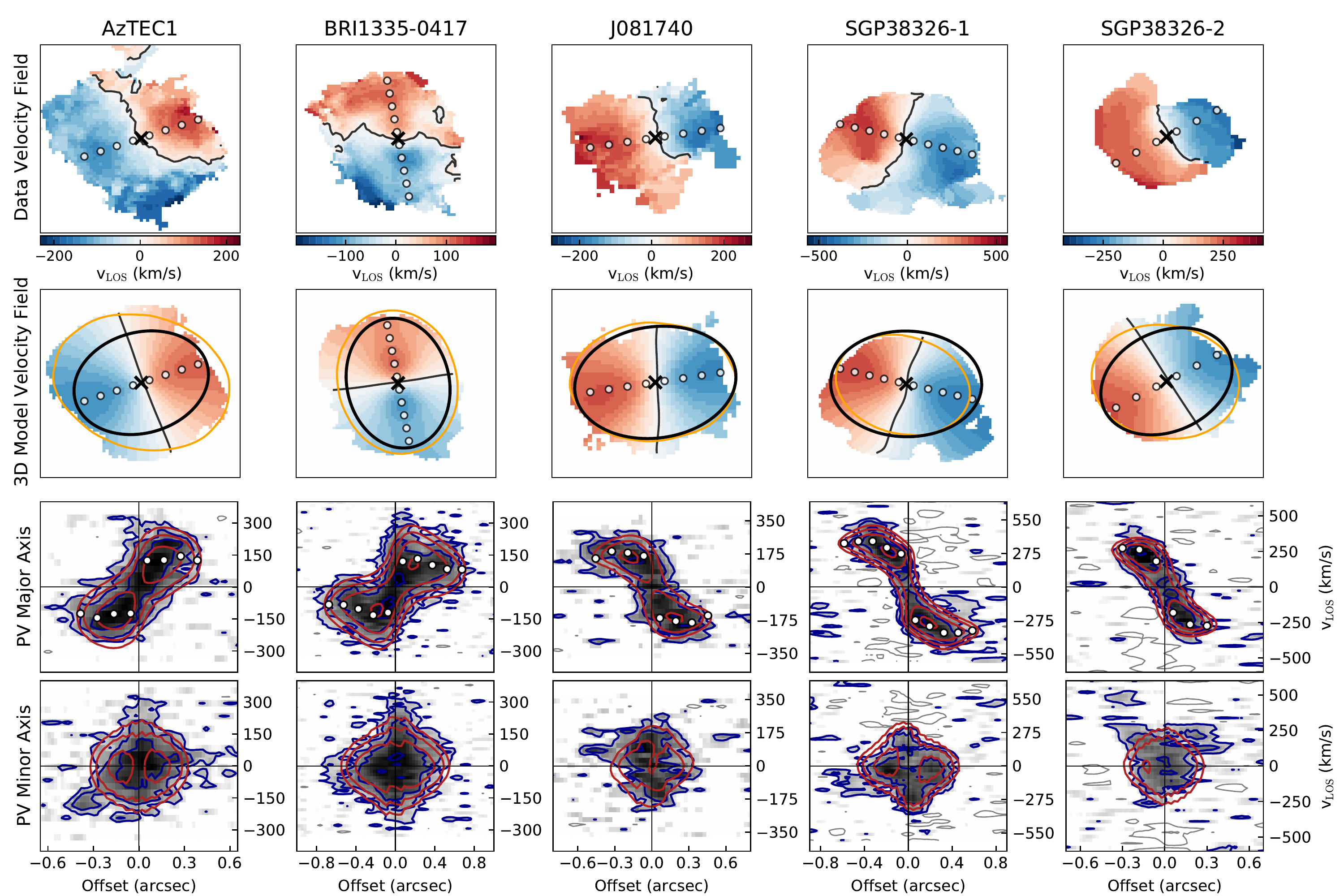}
    \caption{Kinematic best-fit models of our galaxy sample obtained with \texttt{$^{\text{3D}}$BAROLO}. First and second rows: Velocity fields of the data and velocity fields extracted from the best-fit 3D tilted-ring model, respectively. In the second row we show the outer contour of the model obtained by fitting the total intensity map with CANNUBI (in orange, same as in Figure~\ref{fig:data}) and the one obtained with \texttt{$^{\text{3D}}$BAROLO} and used in the kinematic modelling (in black). Third and fourth rows: position-velocity (p-v) diagrams along the major and minor axis, respectively. The data is shown in grayscale with blue and gray contours following positive and negative emission, respectively. The best-fit kinematic model is shown in red contours. The contours follow levels of -2$\sigma_{\mathrm{[CII]}}$ in gray and 2$\sigma_{\mathrm{[CII]}}$, 4$\sigma_{\mathrm{[CII]}}$, 8$\sigma_{\mathrm{[CII]}}$ and 16$\sigma_{\mathrm{[CII]}}$. Our modelled tilted-rings radii are represented by the white markers in the velocity fields following the kinematic PA used in the kinematic models. In the major axis p-v diagram the white markers show the rotation velocities projected along the line of sight.}
    \label{fig:kinmod}
\end{figure*}


\subsubsection{AzTEC1}\label{sec:aztec}
The velocity field and major axis p-v of AzTEC1 show a gradient that the model reproduces reasonably well with a nearly flat velocity distribution. 
It can be seen that the emission in the approaching side of the galaxy flattens more than the receding side, indicating a kinematic asymmetry. As for the minor axis there is a significant emission outside the model contours which tends to broaden the model as \texttt{$^{\text{3D}}$BAROLO} attempts to fit it by increasing the velocity dispersion of the model rotating disc. 

Another evident feature is that the velocity field of the data in Figure~\ref{fig:kinmod} is distorted in the bottom region along the minor axis: it contains a large amount of gas at a velocity around -200 km s$^{-1}$ that bends the iso-velocity contours, including the black contour that traces the central velocity. This region does not affect the rotation curve seen in the p-v of the major axis, however, it can be seen in Figure~\ref{fig:kinmod} as an intense region of emission in the bottom left quadrant of the minor axis p-v that cannot be reproduced with the rotating disc model. Moreover, it can also be seen in the residuals of the channel maps in Figure~\ref{fig:cmA}.

Comparing the two disc geometries shown in the first and second row of Figure~\ref{fig:kinmod}, we can clearly see the misalignment between the morphological PA and the kinematic PA. In the velocity field it is clear that the velocity gradient does not follow the same direction as the elongation of the [CII] total intensity map. Therefore, the morphological PA would not be appropriate to derive the kinematics of the disc as the region with highest line-of-sight velocities would be missed.

All of these features are indications of a rather disturbed kinematics. In Section~\ref{sec:pvsplit} we will see that AzTEC1 is likely an interacting system and we will therefore not use our rotating disc model in the further analysis.

\subsubsection{BRI1335-0417}\label{sec:bri}
The velocity field and major axis p-v of BRI1335-0417 show the velocity rising in the central parts and decreasing and flattening in the external parts. The line profiles are very broad and they extend in the "forbidden" regions located, in this case, in the upper left and, especially, the lower right quadrants of the p-v. These regions of the diagram are characterised as forbidden since a purely rotating disc cannot populate them. Gas emission in these regions signify deviation from circular motions, e.g., show indications of vertical/radial motions due to inflows/outflows and mergers.
The large amount of "forbidden" emission could even be due to the presence of a bar in the central regions of the galaxy, follow-up information on the stellar morphology would be needed to confirm this hypothesis.
Similarly to AzTEC1, the broadness of the emission and the emission in the "forbidden" regions can only be reproduced in our best-fit rotating disc model by increasing the velocity dispersion. This may drive an overestimation of the velocity dispersion at some radii.

We also note a feature close to the centre and most evident in the receding side at high velocities. There is a steep rise of the emission in the centre and an absence of such emission at intermediate/outer radii. This excess of emission in the inner regions can be due to an outflow, with velocities up to 400 km s$^{-1}$. See also the residuals of the channel maps in Figure~\ref{fig:cmB}.
This feature is asymmetric and cannot be reproduced by an axis-symmetric rotating disc model. Therefore, for our fiducial model we prefer to fit the data on the approaching side only to ensure that we are accurately probing the rotation curve.

In the minor axis p-v there is some emission outside of the model, but the most noteworthy feature is an horizontal twist which could further corroborate the need for radial motions in the model.
We tried to fit the kinematics with a free PA, but we found that it does not improve the model significantly.

\subsubsection{J081740}\label{sec:J}
The velocity field and major axis p-v of J081740 show that the emission is less broad in velocity than the previous two cases, indicating that it has lower velocity dispersion. The model contours generally agree with the data contours and reproduces most of the emission without any obvious features. The velocity rises in the centre and flattens out quickly. There is also a tail of emission in the upper left quadrant at lower velocity that may indicate the presence of extraplanar gas \citep[e.g.,][]{marasco19, fraternali02}. However it is uncertain whether it is a real feature or due to background noise.

In the minor axis p-v the model reproduces the data well. There is a bright feature in the upper left quadrant that can also be seen in the 54 km s$^{-1}$ velocity channel in the fifth panel in the residuals of the channel maps in Figure~\ref{fig:cmJ}. This is not part of the rotating disc and could be due to disturbances driven by a minor merger or gas accretion. However, the bulk of the emission is well described by a rotating disc model. This region can also be seen affecting the lower region of the velocity field in the first row of Figure~\ref{fig:kinmod} by distorting the central velocity contour.

\subsubsection{SGP38326-1}\label{sec:SGP1}
The velocity field and major axis p-v of SGP38326-1 show a very clear velocity gradient with the highest line-of-sight velocity of the sample reaching 380 km s$^{-1}$. It has a rising velocity that flattens in the external regions. The kinematic model reproduces most of the bright emission of the galaxy except for a few interesting features.
In the upper left quadrant there is a fainter high receding velocity region that is partially interpreted as high velocity dispersion by the best-fit model, but could be due to a widespread outflow, alike the galactic fountain gas or extraplanar gas seen in local galaxies \citep{fraternali06, boomsma08}.

In the lower right quadrant there is another similar structure that forms a southeast tail in the velocity field that is in the direction of the companion galaxy SGP38326-2 and could be evidence of a tidal interaction between the two galaxies.
Both these features can be seen as residuals in the channel maps in Figure~\ref{fig:cmS1}.
Another interesting feature is that the iso-velocity contours in the velocity field are distorted in the central regions as seen in Figure~\ref{fig:kinmod}.
This could indicate the presence of a warp that we modelled by leaving the PA free in \texttt{$^{\text{3D}}$BAROLO}. We found that the PA ranges from 71$^{\circ}$ in the outermost ring up to 88$^{\circ}$ in the innermost ring. In Figure~\ref{fig:kinmod}, we use the PA of the innermost ring to represent the geometry of the kinematic model. We note that the outermost PA coincides with the morphological PA obtained with \texttt{CANNUBI} as seen in Table~\ref{tab:geometricdata}.
Such warped disc has a very similar kinematics to the model with constant PA, but given the lower residuals, we consider it as our fiducial model.
We note that such distortions could also be due to the presence of radial motions and distinguishing between these two scenarios is difficult \citep{diteodoro21}.

\subsubsection{SGP38326-2}\label{sec:SGP2}
The velocity field and major axis p-v of SGP38326-2 show the velocity rising and flattening, like the other galaxies. The kinematic model reproduces most of the emission and no major external features can be seen.
Although the major axis p-v seems like a very well behaved rotating disc, it is important to note that this galaxy might be undergoing a tidal interaction with SGP38326-1 which makes the kinematic fit more challenging.

In the minor axis, the data contours differ significantly from the model. There are two bright regions in the upper left quadrant that are not reproduced by the rotating disc model. This feature is not exceptional to SGP38326-1, however it could be related to the interaction between SGP38326-1 and SGP38326-2, which is visible in Figure~\ref{fig:sgp}. 

\subsection{Mergers or rotating discs?}\label{sec:pvsplit}

Before we analyse the kinematic models of the galaxies in our sample, we verify the possibility that the velocity gradients that we are observing might be caused by interactions or mergers rather than the presence of a rotating disc. For this, we use the PVsplit diagram, which is a new 3D kinematic classification method to identify the dynamical state of galaxies with the typical resolutions of high-\textit{z} emission-line observations \citep{rizzo22}. Unlike other methods, it is based on the analysis of p-v diagrams instead of the moment 1 and 2 maps. It relies on three parameters: \texttt{Pmajor}, \texttt{PV} and \texttt{PR}. \texttt{Pmajor} quantifies the asymmetries of emission along the major axis p-v diagram relative to the horizontal axis, defining the systemic velocity or redshift. The \texttt{PV} and \texttt{PR} parameters quantify the asymmetries related to the peak of the emission (brightest pixels only) along the major axis p-v diagram. The \texttt{PV} is the asymmetry over the velocity axis and the \texttt{PR} is the asymmetry over the radial axis.

\begin{figure*}
    \centering
    \includegraphics[width=0.95\textwidth]{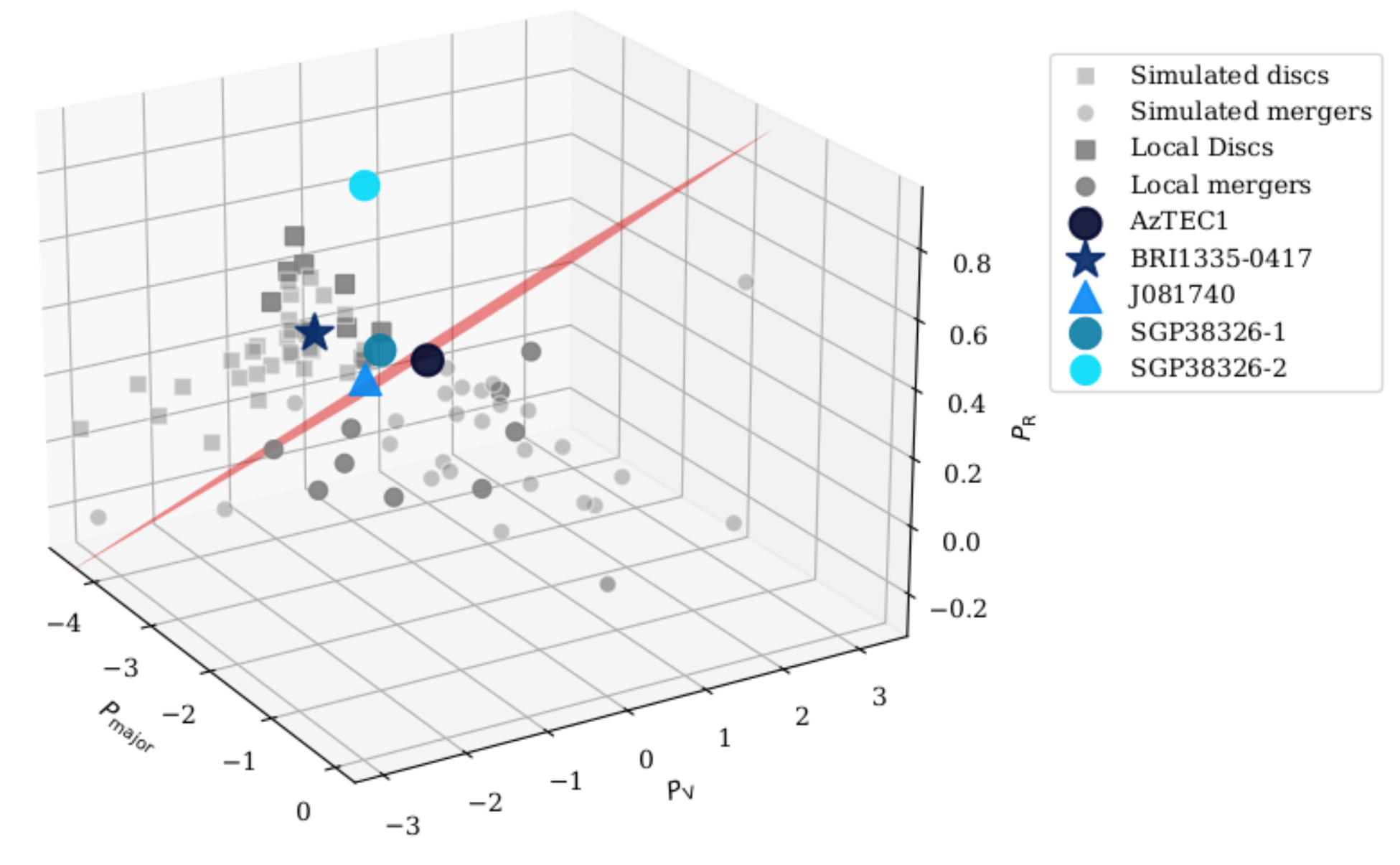}
    \caption{A suitable projection of the position of the galaxies in our sample in the PVsplit diagram, which represents the level of asymmetry in the major-axis p-v. The PVsplit axes are related to asymmetries in the emission in the p-v, where \texttt{Pmajor} is the asymmetry with respect to the origin of the p-v (systemic velocity and galactic centre). The other two parameters define the location of the brightest regions in the p-v with respect to the external line-of-sight velocities (\texttt{PV}) and the outermost radius (\texttt{PR}) used for the kinematic fitting. According to the legend, we show our sample of galaxies in blue markers and categorised in normal star-forming (triangle), quasar host (star) and starburst galaxies (circles). We compare them to simulated galaxies classified as discs (light gray squares) and mergers \citep[light gray circles,][]{rizzo22}. We also show local disc galaxies and mergers from the WHISP survey \citep[dark gray squares and circles, respectively,][]{vanderhulst01}. We choose a projection that is nearly orthogonal to the best-dividing plane shown in red. This plane was obtained by applying the support-vector machine method \citep{cortes95} to the simulated and local galaxies previously classified in \citet{rizzo22}.}\label{fig:pvsplit}
\end{figure*}

We perform this analysis on the galaxies and we plot the results in Figure~\ref{fig:pvsplit}. We compare our galaxies with simulated galaxies known to be mergers or discs taken from \citet{rizzo22}. These are ALMA mock observations of galaxies from the \texttt{SERRA} simulation \citep{pallottini22}. We also show nine randomly selected local disc galaxies from the WHISP survey \citep{vanderhulst01}. From the same survey we show nine mergers, minor and major, that have been classified as mergers by both the presence of companions in the datacubes \citep{swaters02, noordermeer05} and HI morphology \citep{holwerda11}.
Since this is a 3-dimensional analysis, we show the best projection of the PVsplit plane in which we can see the division between the disc galaxies and the mergers.
From Figure~\ref{fig:pvsplit}, it is visible that the two populations occupy different regions in the PVsplit. We highlight this by showing the best-dividing plane for classifying discs and mergers defined by the following equation:

\begin{equation}
    -0.63 P_{\mathrm{major}} - 0.27 P_{\mathrm{V}} + 2.78 P_{\mathrm{R}} - 2.72 = 0.
\end{equation}

The best-dividing plane was obtained by applying the support-vector machine method with a linear kernel, as implemented in the \texttt{sklearn} library \citep{cortes95, scikit-learn}. This method maximises the distance between the two pre-classified samples from each other.

All of our galaxies, except for AzTEC1, are located on the disc side of the distribution. J081740 and SGP38326-1 are close to the boundary but AzTEC1 is located significantly closer to the region populated by merging systems. This is another feature that can support the finding that AzTEC1 could be a merger alongside the large difference in the morphological and kinematic position angles.
In the following analyses of the kinematic properties of the galaxies, we discard the kinematic modelling of AzTEC1 from our sample.

\subsection{Rotation velocities and velocity dispersions}

In Table~\ref{tab:vel}, we display the rotation velocities, velocity dispersions and the V/$\sigma$ ratios for the disc galaxies BRI1335-0417, J081740 SGP38326-1 and SGP38326-2. We report two types of rotation velocities: the maximum rotation velocity (V$_{\mathrm{rot,max}}$) of each rotation curve and the external rotation velocity (V$_{\mathrm{rot,ext}}$). Similarly, we report two values for the velocity dispersion: the mean velocity dispersion across the disc ($\mean{\sigma}$) and the external velocity dispersion ($\sigma_{\mathrm{ext}}$). For the external rotation velocities and velocity dispersions, we average the last two resolution-independent points of the rotation curve and of the velocity dispersion profile, respectively. We note that here we are reporting the rotation velocities, not the circular velocities, thus, these rotation curves should not be used to retrieve dynamical models. These rotation velocities will be corrected for pressure support \citep[see e.g.,][]{iorio17} in the following paper where we will analyse the dynamics in detail.

\begin{table*}
\begin{center}
\caption{Kinematic parameters of the sample: maximum rotation velocity (V$_{\textrm{max}}$), mean velocity dispersion ($\mean{\sigma}$); external rotation velocity (V$_{\textrm{ext}}$); and external velocity dispersion ($\sigma_{\textrm{ext}}$). External is defined as the average of the last two radial points. All of the rotation velocities and velocity dispersions are given in km s$^{-1}$. }\label{tab:vel}
\begin{tabular}{lcccccc}
\hline \hline
ID    & V$_{\textrm{rot,max}}$ & V$_{\textrm{rot,ext}}$ & $\mean{\sigma}$ & $\sigma_{\textrm{ext}}$ & V$_{\textrm{max}}/\mean{\sigma}$ & V$_{\textrm{rot,ext}}/\sigma_{\mathrm{ext}}$ \\ \hline \vspace{2mm}
BRI1335-0417 & 198$^{+17}_{-16}$ & 125$^{+27}_{-26}$ & 75$^{+10}_{-10}$ & 57$^{+11}_{-11}$ & 2.7 $\pm$ 0.4 & 2.2 $\pm$ 0.6 \\ \vspace{2mm}
J081740      & 277$^{+40}_{-38}$ & 249$^{+50}_{-53}$ & 50$^{+13}_{-13}$ & 33$^{+13}_{-13}$ & 5.5 $\pm$ 1.7 & 7.6 $\pm$ 3.4 \\ \vspace{2mm}
SGP38326-1   & 562$^{+24}_{-26}$ & 548$^{+61}_{-71}$ & 58$^{+12}_{-12}$ & 46$^{+11}_{-9}$ & 9.8 $\pm$ 2.0 & 11.8 $\pm$ 2.9 \\ 
SGP38326-2   & 417$^{+51}_{-66}$ & 409$^{+39}_{-48}$ & 49$^{+16}_{-14}$ & 40$^{+13}_{-11}$ & 8.3 $\pm$ 2.8 & 10.2 $\pm$ 3.2 \\ \hline
\end{tabular}
\end{center}
\end{table*}

\begin{figure*}
    \centering
    \includegraphics[width=0.9\textwidth]{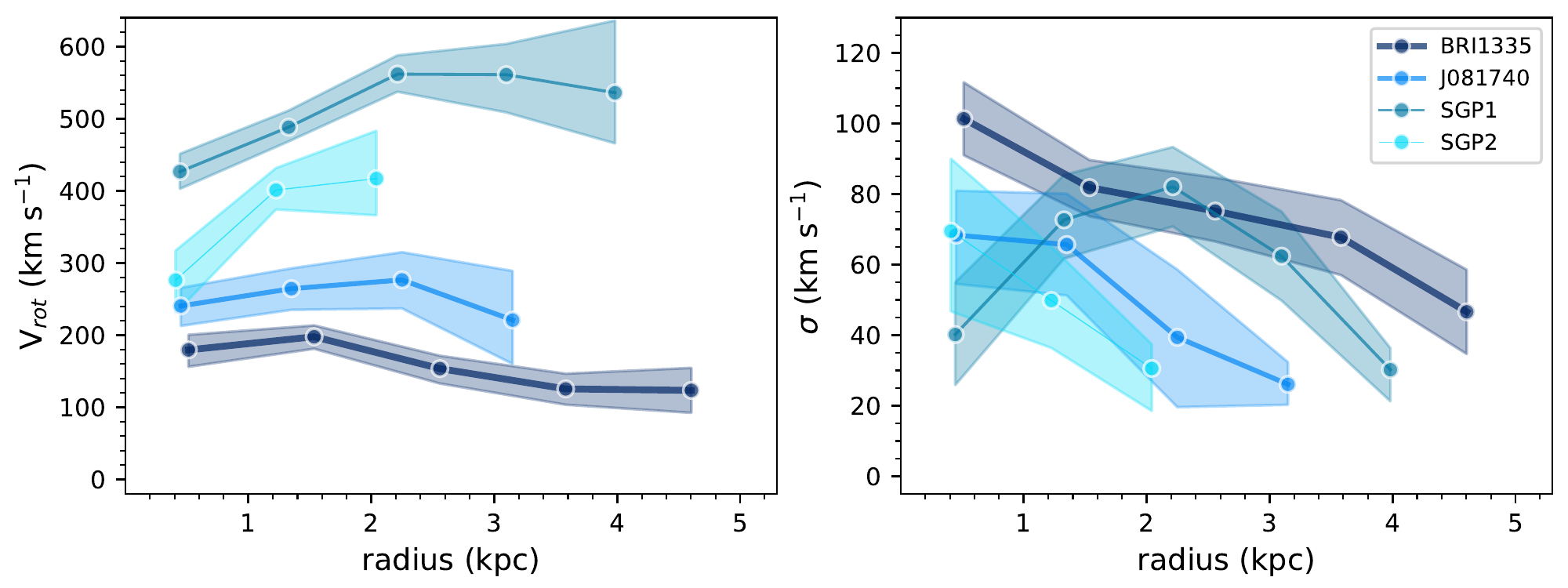}
    \caption{Left panel: Rotation curves of the four discs in our sample. Right panel: Velocity dispersion profiles. The different galaxies are shown in different shades of blue, where the size of the band indicates the estimated errors from the fit. Note that all the radial bins shown here can be considered independent from each other.}\label{fig:rotcur}
\end{figure*}

In Figure~\ref{fig:rotcur}, we show the rotation curves and the velocity dispersion profiles along the radial extent of the galaxies. We see that the galaxies have typically flat rotation curves and declining velocity dispersion profiles. Essentially, all values of velocity dispersions are below 100 km s$^{-1}$.

\begin{figure*}
    \centering
    \includegraphics[width=0.95\textwidth]{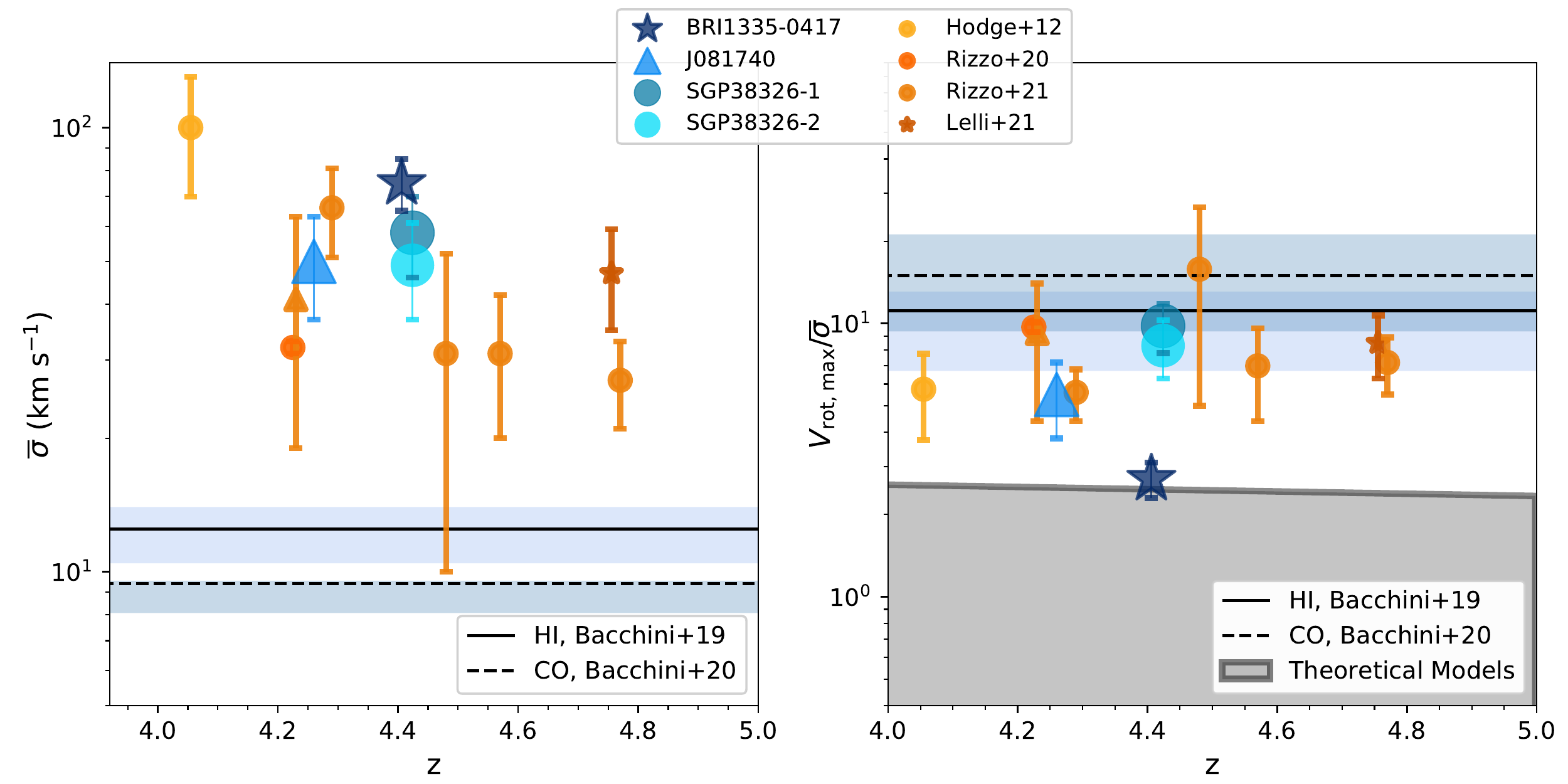}
    \caption{Left Panel: velocity dispersion as a function of redshift. Right Panel: rotation velocity to velocity dispersion ratio as a function of redshift. The galaxies studied in this paper are shown in shades of blue and are categorised in normal star-forming (triangle), AGNs/QSOs host (star) and starburst galaxies (circle). We compare our sample to other well-resolved galaxies in the literature at similar redshift, also analysed with a 3D methodology and observed in [CII], or CO(2-1) in the case of GN20 \citep{hodge12}, shown in different shades of orange \citep{hodge12, rizzo20, rizzo21, lelli21}. We show average velocity dispersions and V/$\sigma$ of spiral galaxies in the Local Universe measured in HI and CO, as solid and dashed lines respectively \citep{bacchini19, bacchini20}. The blue shading marks the 84th and 16th percentiles of the HI and CO measurements. On the right panel, we show the expectations of galaxy formation models for this redshift in gray shading \citep{dekel14, zolotov15, hayward17, pillepich19}.}\label{fig:sigma}
\end{figure*}

To analyse these results, we choose to focus on two values in Figure~\ref{fig:sigma}: $\mean{\sigma}$ and V$_{\mathrm{rot,max}}/\mean{\sigma}$. The average velocity dispersion ($\mean{\sigma}$) provides insight into the overall turbulence across the disc and the ratio of maximum rotation velocity and the average dispersion (V$_{\mathrm{rot,max}}/\mean{\sigma}$) can tell us about how much dynamical support is due to ordered or turbulent motions.
In Figure~\ref{fig:sigma}, we show the trends of the dispersion velocities of this sample in comparison with an expanded sample of other galaxies in the literature at a similar redshift, observed in [CII] or CO, well-resolved and analysed with a 3D kinematic methodology.
We remind the reader that a 3D methodology is necessary in order to fully correct for beam smearing effects in low-resolution data.
We see that our sample is similar to other galaxies with dynamically cold discs at similar redshift \citep{hodge12, rizzo20, rizzo21, lelli21}.
We note that there are other works at this redshift range that we do not compare in Figure~\ref{fig:sigma} due to poor angular resolution \citep[e.g.,][]{jones21, fraternali21}, due to a 2D kinematic analysis \citep[e.g.,][]{carniani13, jones17}.
Additionally, other works outside our redshift range have analysed the kinematics of other sources with indications of rotating discs both in star-forming galaxies \citep{smit18, herreracamus22, tokuoka22} and in quasar host galaxies \citep{neeleman21, pensabene20, yue21, shao22}.

Additionally, we can see a significant evolution in velocity dispersions when compared to HI and CO measurements of Local Universe spiral galaxies \citep[solid and dashed lines in Figure~\ref{fig:sigma}]{bacchini19, bacchini20}.
We choose to show the data from \citet{bacchini19} and \citet{bacchini20} given that they are also analysed with \texttt{$^{\text{3D}}$BAROLO}, although they are in agreement with other studies at low redshift, e.g., \citet{narayan02, tamburro09}.
Furthermore, this is in line with the trends seem at lower redshift \citep{uebler19}, but our values for the velocity dispersion are typically lower than found in some intermediate redshift galaxies observed in cold gas \citep{kaasinen20} or ionised gas \citep{turner17}.
We note that the [CII] line emission traces different gas phases \citep{wolfire22}. However, the velocity dispersions of these three tracers are observed to be similar in nearby galaxies, in particular [CII] agrees well with the CO distribution and linewidths \citep{deblok16}.

In Figure~\ref{fig:sigma}, we see that most galaxies of our sample and the extended sample have dynamically cold discs, comparable to the ones found in the Local Universe and systematically above the theoretical expectations \citep[e.g.,][]{dekel14, zolotov15, hayward17, pillepich19}, although it is clear that our high V/$\sigma$ values are largely driven by the fast rotations. The high V/$\sigma$ and relatively low $\sigma$ are consistent, instead, with recent predictions obtained using zoom-in simulations on \textit{z} $\gtrsim 6$ main-sequence galaxies \citep[e.g.,][]{kohandel20, tamfal22}. The only exception is the quasar host galaxy BRI1335-0417, which is the only galaxy in marginal agreement with the model expectations.
In the following Section, we further discuss the interpretations of the kinematic models shown here.

\section{Discussion}\label{sec:discussion}

\subsection{Gas kinematics at \textit{z} $\sim$ 4.5}

In the previous Section, we have presented the ALMA [CII] 158 $\mu$m datasets of five galaxies at \textit{z} $\sim$ 4.5 and the results of the kinematic models obtained with \texttt{$^{\text{3D}}$BAROLO}. The sample is very diverse: AzTEC1 is a starburst galaxy, BRI1335-0417 is a starburst and quasar host galaxy, J081740 is a normal star-forming galaxy and SGP38326-1 and SGP38326-2 are starburst galaxies part of a group.
In this Section, we discuss the comparison with previous works, the reliability of our results as well as how do these galaxies fit into the context of other kinematically cold discs at \textit{z} $\sim$ 4.5.

\paragraph*{AzTEC1}
Previous studies of AzTEC1 found a velocity gradient in the velocity fields as traced by CO(J=4-3), [CII] 158 $\mu$m and [NII] 205 $\mu$m emission. With \texttt{GalPak3D} \citep{bouche15}, it was found that that the velocity gradient in the [CII] 158 $\mu$m is consistent with a rotating disc model with a V/$\sigma \sim 2.5$ \citep{tadaki19, tadaki18}.
However, many off-centre clumps and non-corotating regions have been found in AzTEC1 in the gas and dust emission \citep{iono16, tadaki20}, including the clump that we mention in Section~\ref{sec:aztec}. The latter was identified by \citet{tadaki20} as a non-corotating inflow of gas, likely a gas rich minor merger.
When also taking into account the misalignment of the morphological and kinematic PA identified in Section~\ref{sec:disc} and the position in the PVsplit diagram in Figure~\ref{fig:pvsplit}, we find that AzTEC1 is likely a spatially unresolved interacting system or merger and thus the velocity difference between the two interacting galaxies can contribute to the observed velocity gradient. For this reason we do not show values of V and $\sigma$ in Figure~\ref{fig:rotcur} and Table~\ref{tab:vel} as it may not reliably trace the dynamical information of an underlying rotating disc.
The merger can also be an explanation for the high [CII] flux and SFR. Given that the dataset analysed in this work has an integration time of only 30 min with ALMA, it is possible that deeper observations at higher angular resolution can solve this issue.
Furthermore, future observations with the James Webb Space Telescope (JWST) could also provide insights into the nature of AzTEC1 and, perhaps, confirm the merger scenario.

\paragraph*{BRI1335-0417}
As mentioned in Section~\ref{sec:data}, a recent study of BRI1335-0417 analysing the CO(J=7-6) emission found evidence of a bipolar central outflow up to $\sim$600 km s$^{-1}$ that can be associated with the quasar \citep{lu18}. In the [CII] 158 $\mu$m emission we find a feature that can be a match to the feature seen in \citet{lu18}. The presence of this high-velocity feature in the centre of the galaxy may be driving an overestimation of the velocity dispersion at inner radii, as seen in the velocity dispersion profile in Figure~\ref{fig:rotcur}. Regardless, the average velocity dispersion across the disc is 75 $\pm$ 10 km s$^{-1}$ in spite of the intense SFR of 5$\times 10^3$M$_{\odot}$ yr$^{-1}$. The rotation velocity was found to be 198$^{+17}_{-16}$ km s$^{-1}$, consistent with the dynamical model obtained by \citet{tsukui21} using \texttt{KinMS} \citep{davis13}. Therefore, BRI1335-0417 has a V/$\sigma$ of 2.7. Making it the rotationally supported disc with the lowest V/$\sigma$ among our sample and the other galaxies at similar redshift shown in Figure~\ref{fig:sigma}. Despite the low V/$\sigma$ it shows a rather undisturbed disc in the PVsplit analysis.

\paragraph*{J081740}
Previous kinematic analysis of J081740 found that it has a rotation velocity of 272$^{+52}_{-13}$ km $s^{-1}$ and a velocity dispersion of 80$^{+4}_{-11}$ km $s^{-1}$ obtained with \texttt{Qubefit} by \citet{neeleman20} and a rotation velocity of $250 \pm 15$ km $s^{-1}$ and a velocity dispersion of $34 \pm 11$ km $s^{-1}$ also obtained with \texttt{$^{\text{3D}}$BAROLO} by \citep{jones21}. In this paper, we find a maximum rotation velocity of 277$^{+40}_{-38}$ km $s^{-1}$ and average velocity dispersion across the disc of 50$^{+13}_{-13}$ km $s^{-1}$. Therefore, our results are in between the previous two estimates, although the velocity dispersion is closer to the results in \citet{jones21}. There is a general compatibility within 1-$\sigma$ errors, except for the velocity dispersion found in \citet{neeleman20}. The differences in the estimations are likely due to the differences in the imaging of the datacube, assumptions between \texttt{Qubefit} and \texttt{$^{\text{3D}}$BAROLO} as well as the different parameters used for the masking and geometry.
We estimate a V/$\sigma$ of 5.5, in line with other highly-resolved disc galaxies and higher than the models predict. This is an interesting result, as J081740 is the galaxy with the lowest star formation rate among the galaxies analysed at similar redshift, and likely the closest to the star-forming main sequence. This can only be confirmed with an estimate of the stellar mass and could suggest that also main-sequence galaxies at \textit{z} $\sim 4.5$ host cold rotating discs. As the other galaxies in our sample, we also see non-rotating features that can indicate a minor merger or inflow/outflow as discussed in Section~\ref{sec:kin}.

\paragraph*{SGP38326-1 and SGP38326-2}
In \citet{oteo16} it was found that the CO emission (traced by CO(J=4-3) and CO(J=5-4)) forms a bridge in between the two main galaxies, however, no [CII] bridge was found. The dataset we used is considerably deeper and, although, we cannot observe a bridge in the [CII] emission of the two galaxies, we can see a non-corotating feature in SGP38326-1 and SGP38326-2 that could be evidence of a [CII] tidal tail or connection between the galaxies, adding to the evidence of an interaction.
%
Regardless of these features, we identify two rotating discs with maximum rotation velocities of 562 $^{+24}_{-26}$ km s$^{-1}$ and 417$^{+51}_{-60}$ km s$^{-1}$ and average velocity dispersions of 58 $^{+12}_{-12}$ km s$^{-1}$ and 49$^{+16}_{-14}$ km s$^{-1}$ for SGP38326-1 and SGP38326-2, respectively. This means they have with V$/\sigma$ of 9.8 and 8.3, similar to the other rotating discs at \textit{z} $\sim$ 4.5.
The newly discovered [CII] emission of SGP38326-3 and SGP38326-4 shown in Figure~\ref{fig:sgp} indicates that this might be a rich group, possibly the early stages of a proto-cluster. Deeper observations and over a larger bandwidth could find more companions.

\vspace{6mm}
The general picture that we obtain is that rotating gas discs with V$/\sigma \gtrsim 5$ are common among many different types of galaxies at \textit{z} $\sim$ 4.5.
This is in line with early JWST results that show significantly higher disc fractions than previously expected at high-\textit{z} and that disc galaxies are the dominant morphology up to \textit{z} = 6 \citep{ferreira22, robertson23}.
In all five galaxies in our sample, there are non-corotating regions which can be indications of gas accretion or outflows.
The average velocity dispersions across the discs of our sample range between 50 km s$^{-1}$ and 75 km s$^{-1}$. BRI1335-0417 has a slightly higher velocity dispersion than other galaxies, albeit compatible within 1-$\sigma$ errors, which leads to a V$/\sigma$ marginally compatible with theoretical expectations \citep{pillepich19}.
The different V/$\sigma$ value can be due to the fact that we are comparing submillimetre galaxies \citep{rizzo20, lelli21, rizzo21} with a quasar host galaxy, in the latter high tubulent motions can be driven by the AGN feedback.
BRI1335-0417 shows extensive emission in the "forbidden" regions of the major axis p-v. These features are usually dynamical evidence of the presence of bars in local galaxies and could be later confirmed by observations of the stellar component of the galaxies with JWST, but they can also be evidence of strong non-circular (likely outflow) motions.
We also observe some distortions in the iso-velocities of SGP38326-1 that can indicate radial motions or a warp.

Finally, we stress the challenges of deriving reliable kinematic parameters in high-\textit{z} galaxies. The typical limitations in SNR and spatial resolution of observations of high-\textit{z} galaxies can bias the interpretation of the kinematics. In \citet{rizzo22}, it was shown using mock data obtained from hydrodynamical simulations that, depending on the methodology, discs can be completely misclassified as mergers or dispersion dominated systems, while up to 50\% of mergers can be misclassified as discs.
Therefore, the presence of velocity gradients is not sufficient to classify a system as a disc.



\subsection{Disc thickness}\label{sec:discthick}

Finally, in this Section we discuss the fact that all the galaxies analysed here have inclinations close to $\sim$42$^{\circ}$, well below 60$^{\circ}$ which is the expected median inclination of a sample composed of galaxies with random inclinations \citep[see A.2]{romanowsky12}. A possible bias that could favour the selection of low-inclination galaxies in line-emission observations is the fact that high-inclination galaxies have their emission spread out over a wider velocity range when compared to low-inclination galaxies with the same line flux \citep{kohandel19}. This bias is expected to be important in the case of faint sources close to the detection limit and might indeed influence J081740 and SGP36326-2 but not SGP3626-1 and BRI1335-0417 given that the latter are on the bright end of the [CII] flux distribution of non-lensed sources at \textit{z} $\sim 4$. Naturally, due to the small sample size, the galaxies may just coincidentally have similar inclinations, however, it could also be that we are systematically underestimating their inclinations due to our thin disc assumption.

To investigate how important the effect of disc thickness is in our kinematic modelling, we attempted a direct determination of the disc thickness with \texttt{CANNUBI}.
As discussed in Section~\ref{sec:CANNUBI}, for thick discs the best approach is to run \texttt{CANNUBI} over the full data cube to better discern the degeneracy between inclination and disc thickness. So we run \texttt{CANNUBI} on the galaxies keeping the PA fixed to the morphological PA and leaving the inclination and disc thickness (constant with radius) free. Our results are the following: for BRI1335-0417 we obtain a disc thickness of 1.6 $\pm$ 0.1 kpc with a disc inclination of $58^{+8}_{-5}$ degrees; for J081740 we obtain a disc thickness of 1.3 $\pm$ 0.1 kpc with a disc inclination of $69^{+11}_{-7}$ degrees; for SGP38326-1 we obtain a disc thickness of 1.2 $\pm$ 0.2 kpc with a disc inclination of $56^{+13}_{-9}$ degrees; for SGP38326-2 we obtain a disc thickness of 0.5 $\pm$ 0.1 kpc with a disc inclination of $44^{+3}_{-4}$ degrees. 
Thus, considering the thickness, the inclinations are slightly increased and closer to 60$^{\circ}$ while the disc thickness are of the order of 1 kpc for all galaxies. Interestingly, the  thickness of 1 kpc found for our sample are consistent with the typical scale heights of the thick discs in the Local universe, such as the Milky Way \citep{li17, bland-hawthorn16A}. This might indicate that thick stellar discs can be formed locally from the thick gas disc in the early stages of galaxy formation \citep{brook04}. 
However, the reliability of these results needs to be checked in particular using mock data derived from simulated galaxies, an investigation that we leave to future works.

Despite the open questions outlined above, it is important to quantify the impact of thickness on the kinematics. The inclinations retrieved simultaneously with the disc thickness represent an increase of 15 degrees in the average inclination of the sample. We have then rerun \texttt{$^{\text{3D}}$BAROLO} fixing the inclinations and thicknesses to the values reported above for the individual galaxies and obtained new best-fit rotations and velocity dispersions. This resulted in an average decrease of 10\% in the maximum rotation velocity of the galaxies, while the changes in the velocity dispersion are within the errors and the rotation curves retained the same shape.

An important caveat here is that the ring by ring approach used in \texttt{$^{\text{3D}}$BAROLO} is not ideal to perform fits where the thickness of the discs is larger than the spatial resolution of the observations, as it happens in particular in the case of BRI1335-0417. However, even considering a velocity correction based only on the change in inclination with a simple ratio between the $\sin$ of the inclinations, the highest impact would be to reduce the rotation velocity and V/$\sigma$ of J081740 by 20\% and less than that for the other galaxies.

Overall our general results that the discs of these galaxies are cold and the V/$\sigma$ values are significantly larger than model predictions remain unchanged. Finally, we note that in Figure~\ref{fig:rotcur}, we show only errors related to the kinematic fitting. However, for the upcoming dynamical modelling we will also take into account the systematic uncertainties due to the impact of the thickness on the inclination.

\section{Conclusions}\label{sec:conclusions}
The purpose of this study was to further understand the dynamics of discs in the early Universe. We presented high-resolution ALMA observations, at scales in the range of 0.9 to 1.2 kpc, of the [CII] 158 $\mu$m emission line of five galaxies at \textit{z} $\sim$ 4.5.
We derived robust kinematic parameters, namely rotation velocities and velocity dispersions at various galactocentric radii for each of the four galaxies using \texttt{$^{\text{3D}}$BAROLO}. This methodology properly accounts for beam smearing effects.
We determined the disc geometry of the galaxies including an attempt at directly measuring their disc thickness.
The main results of this paper are summarised below:

\begin{enumerate}
    \item Of the five galaxies we find that four of them have rotationally supported discs (BRI1335-0417, J081740, SGP38326-1 and SGP38326-2) and one of them is likely a close interaction or a merger that is spatially unresolved (AzTEC1). Together with other high-resolution galaxy observations at the same redshift range, we confirm that discs are already in place at \textit{z} $\sim$ 4.
    
    \item The rotating disc galaxies at \textit{z} $\sim$ 4.5 have flat rotation curves with maximum rotation velocities ranging between 198 and 562 km s$^{-1}$ and velocity dispersions averaged across the discs ranging between 50 and 75 km s$^{-1}$. The ratios of rotation-to-random motions (V/$\sigma$) are 2.7, 5.5, 9.8 and 8.3 for BRI1335-0417, J081740, SGP38326-1 and SGP38326-2, respectively. These V/$\sigma$ values are significantly outside the predictions of current models, with the exception of BRI1335-0417 that is in marginal agreement with expectations.
    
    \item In spite of the presence of regularly rotating discs, we found that non-circular motions are ubiquitous among the galaxies in our sample. These features can be evidence of interactions or gas inflow/outflow. In BRI1335-0417, an outflow is likely responsible for the high [CII] velocities observed close to the galactic centre, while the interpretation of other features require more investigation.
    \item We discover [CII] emission in a third and fourth galaxy in the SGP38326 system and we find evidence that the two main galaxies might be tidally interacting, as shown in [CII], which was not clear from previous works with shallower and lower resolution data.
\end{enumerate}

Although it is hard to create a statistical picture of the kinematics at high-\textit{z} due to the lack of highly resolved observations of large samples of galaxies, this study shows that rotating discs can be common occurrences at \textit{z} $\sim$ 4.5 in spite of the existence of several complex features in their kinematics. 
For our kinematic analysis, we used the standard approach of assuming razor-thin discs or, more realistically, a disc thickness that is significantly lower than the resolution of the data. However, we also made the first attempt towards an estimate of their thickness and found values of the order of 1 kpc. 


In a follow-up study we will cover the origin of turbulence (e.g., due to stellar feedback or gravitational instabilities), the mass decomposition of the rotation curves obtained in this work and investigate the location of these galaxies in scaling relations.

\section*{Acknowledgements}%
We thank the anonymous referee for their careful and valuable feedback. FRO and FF acknowledge support from The Dutch Research Council (NWO) through the Klein-1 Grant code OCEN2.KLEIN.088. FR acknowledges support from the European Union’s Horizon 2020 research and innovation program under the Marie Sklodowska-Curie grant agreement No. 847523 ‘INTERACTIONS’ and the Cosmic Dawn Center that is funded by the Danish National Research Foundation under grant No. 140. We acknowledge extensive assistance from Allegro, the European ALMA Regional Center node in the Netherlands. The Joint ALMA Observatory is operated by ESO, AUI/NRAO and NAOJ. This research has made use of the NASA/IPAC Extragalactic Database (NED), which is operated by the Jet Propulsion Laboratory, California Institute of Technology, under contract with the National Aeronautics and Space Administration.
This work makes use of the following ALMA data: ADS/JAO.ALMA\#2015.1.00330.S, ADS/JAO.ALMA\#2017.1.00127.S, \#2017.1.00394.S, \#2017.1.01052.S, \#2018.1.00001.S. ALMA is a partnership of ESO (representing its member states), NSF (USA) and NINS (Japan), together with NRC (Canada), MOST and ASIAA (Taiwan), and KASI (Republic of Korea), in cooperation with the Republic of Chile.



\section*{Data Availability}
The data underlying this article are public and available in the ALMA Science Archive, at \url{https://almascience.nrao.edu/asax/}. The clean datacubes, parameter files and outputs of \texttt{$^{\text{3D}}$BAROLO} are available in the Zenodo (DOI:10.5281/zenodo.7543961).
\texttt{$^{\text{3D}}$BAROLO} is available at \url{http://editeodoro.github.io/Bbarolo/}. \texttt{CANNUBI} is available at \url{https://www.filippofraternali.com/cannubi}.

\bibliographystyle{mnras}
\bibliography{references.bib}

\begin{thebibliography}{}
\makeatletter
\relax
\def\mn@urlcharsother{\let\do\@makeother \do\$\do\&\do\#\do\^\do\_\do\%\do\~}
\def\mn@doi{\begingroup\mn@urlcharsother \@ifnextchar [ {\mn@doi@}
  {\mn@doi@[]}}
\def\mn@doi@[#1]#2{\def\@tempa{#1}\ifx\@tempa\@empty \href
  {http://dx.doi.org/#2} {doi:#2}\else \href {http://dx.doi.org/#2} {#1}\fi
  \endgroup}
\def\mn@eprint#1#2{\mn@eprint@#1:#2::\@nil}
\def\mn@eprint@arXiv#1{\href {http://arxiv.org/abs/#1} {{\tt arXiv:#1}}}
\def\mn@eprint@dblp#1{\href {http://dblp.uni-trier.de/rec/bibtex/#1.xml}
  {dblp:#1}}
\def\mn@eprint@#1:#2:#3:#4\@nil{\def\@tempa {#1}\def\@tempb {#2}\def\@tempc
  {#3}\ifx \@tempc \@empty \let \@tempc \@tempb \let \@tempb \@tempa \fi \ifx
  \@tempb \@empty \def\@tempb {arXiv}\fi \@ifundefined
  {mn@eprint@\@tempb}{\@tempb:\@tempc}{\expandafter \expandafter \csname
  mn@eprint@\@tempb\endcsname \expandafter{\@tempc}}}

\bibitem[\protect\citeauthoryear{{Bacchini}, {Fraternali}, {Iorio}  \&
  {Pezzulli}}{{Bacchini} et~al.}{2019}]{bacchini19}
{Bacchini} C.,  {Fraternali} F.,  {Iorio} G.,   {Pezzulli} G.,  2019, \mn@doi
  [\aap] {10.1051/0004-6361/201834382}, \href
  {https://ui.adsabs.harvard.edu/abs/2019A&A...622A..64B} {622, A64}

\bibitem[\protect\citeauthoryear{{Bacchini}, {Fraternali}, {Iorio}, {Pezzulli},
  {Marasco}  \& {Nipoti}}{{Bacchini} et~al.}{2020}]{bacchini20}
{Bacchini} C.,  {Fraternali} F.,  {Iorio} G.,  {Pezzulli} G.,  {Marasco} A.,
  {Nipoti} C.,  2020, \mn@doi [\aap] {10.1051/0004-6361/202038223}, \href
  {https://ui.adsabs.harvard.edu/abs/2020A&A...641A..70B} {641, A70}

\bibitem[\protect\citeauthoryear{{Begeman}}{{Begeman}}{1987}]{begeman87}
{Begeman} K.~G.,  1987, PhD thesis, University of Groningen

\bibitem[\protect\citeauthoryear{{Biernacki} \& {Teyssier}}{{Biernacki} \&
  {Teyssier}}{2018}]{biernacki18}
{Biernacki} P.,  {Teyssier} R.,  2018, \mn@doi [\mnras] {10.1093/mnras/sty216},
  \href {https://ui.adsabs.harvard.edu/abs/2018MNRAS.475.5688B} {475, 5688}

\bibitem[\protect\citeauthoryear{{Bland-Hawthorn} \&
  {Gerhard}}{{Bland-Hawthorn} \& {Gerhard}}{2016}]{bland-hawthorn16A}
{Bland-Hawthorn} J.,  {Gerhard} O.,  2016, \mn@doi [\araa]
  {10.1146/annurev-astro-081915-023441}, \href
  {https://ui.adsabs.harvard.edu/abs/2016ARA&A..54..529B} {54, 529}

\bibitem[\protect\citeauthoryear{{Boomsma}, {Oosterloo}, {Fraternali}, {van der
  Hulst}  \& {Sancisi}}{{Boomsma} et~al.}{2008}]{boomsma08}
{Boomsma} R.,  {Oosterloo} T.~A.,  {Fraternali} F.,  {van der Hulst} J.~M.,
  {Sancisi} R.,  2008, \mn@doi [\aap] {10.1051/0004-6361:200810120}, \href
  {https://ui.adsabs.harvard.edu/abs/2008A&A...490..555B} {490, 555}

\bibitem[\protect\citeauthoryear{{Bosma}}{{Bosma}}{1978}]{bosma78}
{Bosma} A.,  1978, PhD thesis, University of Groningen

\bibitem[\protect\citeauthoryear{{Bouch{\'e}}, {Carfantan}, {Schroetter},
  {Michel-Dansac}  \& {Contini}}{{Bouch{\'e}} et~al.}{2015}]{bouche15}
{Bouch{\'e}} N.,  {Carfantan} H.,  {Schroetter} I.,  {Michel-Dansac} L.,
  {Contini} T.,  2015, \mn@doi [\aj] {10.1088/0004-6256/150/3/92}, \href
  {https://ui.adsabs.harvard.edu/abs/2015AJ....150...92B} {150, 92}

\bibitem[\protect\citeauthoryear{Briggs}{Briggs}{1995}]{briggs95}
Briggs D.~S.,  1995, PhD thesis, The New Mexico Institute of Mining and
  Technology

\bibitem[\protect\citeauthoryear{{Brook}, {Kawata}, {Gibson}  \&
  {Freeman}}{{Brook} et~al.}{2004}]{brook04}
{Brook} C.~B.,  {Kawata} D.,  {Gibson} B.~K.,   {Freeman} K.~C.,  2004, \mn@doi
  [\apj] {10.1086/422709}, \href
  {https://ui.adsabs.harvard.edu/abs/2004ApJ...612..894B} {612, 894}

\bibitem[\protect\citeauthoryear{{Capak} et~al.,}{{Capak}
  et~al.}{2015}]{capak15}
{Capak} P.~L.,  et~al., 2015, \mn@doi [\nat] {10.1038/nature14500}, \href
  {https://ui.adsabs.harvard.edu/abs/2015Natur.522..455C} {522, 455}

\bibitem[\protect\citeauthoryear{{Carilli} \& {Walter}}{{Carilli} \&
  {Walter}}{2013}]{carilli&walter13}
{Carilli} C.~L.,  {Walter} F.,  2013, \mn@doi [\araa]
  {10.1146/annurev-astro-082812-140953}, \href
  {https://ui.adsabs.harvard.edu/abs/2013ARA&A..51..105C} {51, 105}

\bibitem[\protect\citeauthoryear{{Carniani} et~al.,}{{Carniani}
  et~al.}{2013}]{carniani13}
{Carniani} S.,  et~al., 2013, \mn@doi [\aap] {10.1051/0004-6361/201322320},
  \href {https://ui.adsabs.harvard.edu/abs/2013A&A...559A..29C} {559, A29}

\bibitem[\protect\citeauthoryear{{Chabrier}}{{Chabrier}}{2003}]{chabrier03}
{Chabrier} G.,  2003, \mn@doi [\pasp] {10.1086/376392}, \href
  {https://ui.adsabs.harvard.edu/abs/2003PASP..115..763C} {115, 763}

\bibitem[\protect\citeauthoryear{Cortes \& Vapnik}{Cortes \&
  Vapnik}{1995}]{cortes95}
Cortes C.,  Vapnik V.,  1995, Machine Learning, 20, 273

\bibitem[\protect\citeauthoryear{{Courteau}}{{Courteau}}{1997}]{courteau97}
{Courteau} S.,  1997, \mn@doi [\aj] {10.1086/118656}, \href
  {https://ui.adsabs.harvard.edu/abs/1997AJ....114.2402C} {114, 2402}

\bibitem[\protect\citeauthoryear{{Davis}, {Bureau}, {Cappellari}, {Sarzi}  \&
  {Blitz}}{{Davis} et~al.}{2013}]{davis13}
{Davis} T.~A.,  {Bureau} M.,  {Cappellari} M.,  {Sarzi} M.,   {Blitz} L.,
  2013, \mn@doi [\nat] {10.1038/nature11819}, \href
  {https://ui.adsabs.harvard.edu/abs/2013Natur.494..328D} {494, 328}

\bibitem[\protect\citeauthoryear{{Davis}, {Zabel}  \& {Dawson}}{{Davis}
  et~al.}{2020}]{davis20}
{Davis} T.~A.,  {Zabel} N.,   {Dawson} J.~M.,  2020, {KinMS: Three-dimensional
  kinematic modelling of arbitrary gas distributions}, Astrophysics Source Code
  Library, record ascl:2006.003 (\mn@eprint {ascl} {2006.003})

\bibitem[\protect\citeauthoryear{{Dekel} \& {Burkert}}{{Dekel} \&
  {Burkert}}{2014}]{dekel14}
{Dekel} A.,  {Burkert} A.,  2014, \mn@doi [\mnras] {10.1093/mnras/stt2331},
  \href {https://ui.adsabs.harvard.edu/abs/2014MNRAS.438.1870D} {438, 1870}

\bibitem[\protect\citeauthoryear{{Dekel}, {Sari}  \& {Ceverino}}{{Dekel}
  et~al.}{2009}]{dekel09}
{Dekel} A.,  {Sari} R.,   {Ceverino} D.,  2009, \mn@doi [\apj]
  {10.1088/0004-637X/703/1/785}, \href
  {https://ui.adsabs.harvard.edu/abs/2009ApJ...703..785D} {703, 785}

\bibitem[\protect\citeauthoryear{{Dekel}, {Ginzburg}, {Jiang}, {Freundlich},
  {Lapiner}, {Ceverino}  \& {Primack}}{{Dekel} et~al.}{2020}]{dekel20}
{Dekel} A.,  {Ginzburg} O.,  {Jiang} F.,  {Freundlich} J.,  {Lapiner} S.,
  {Ceverino} D.,   {Primack} J.,  2020, \mn@doi [\mnras]
  {10.1093/mnras/staa470}, \href
  {https://ui.adsabs.harvard.edu/abs/2020MNRAS.493.4126D} {493, 4126}

\bibitem[\protect\citeauthoryear{{Di Teodoro} \& {Fraternali}}{{Di Teodoro} \&
  {Fraternali}}{2015}]{diteodoro15}
{Di Teodoro} E.~M.,  {Fraternali} F.,  2015, \mn@doi [\mnras]
  {10.1093/mnras/stv1213}, \href
  {https://ui.adsabs.harvard.edu/abs/2015MNRAS.451.3021D} {451, 3021}

\bibitem[\protect\citeauthoryear{{Di Teodoro} \& {Peek}}{{Di Teodoro} \&
  {Peek}}{2021}]{diteodoro21}
{Di Teodoro} E.~M.,  {Peek} J.~E.~G.,  2021, \mn@doi [\apj]
  {10.3847/1538-4357/ac2cbd}, \href
  {https://ui.adsabs.harvard.edu/abs/2021ApJ...923..220D} {923, 220}

\bibitem[\protect\citeauthoryear{{Ferreira} et~al.,}{{Ferreira}
  et~al.}{2022}]{ferreira22}
{Ferreira} L.,  et~al., 2022, \mn@doi [\apjl] {10.3847/2041-8213/ac947c}, \href
  {https://ui.adsabs.harvard.edu/abs/2022ApJ...938L...2F} {938, L2}

\bibitem[\protect\citeauthoryear{{Foreman-Mackey}, {Hogg}, {Lang}  \&
  {Goodman}}{{Foreman-Mackey} et~al.}{2013}]{mackey13}
{Foreman-Mackey} D.,  {Hogg} D.~W.,  {Lang} D.,   {Goodman} J.,  2013, \mn@doi
  [\pasp] {10.1086/670067}, \href
  {https://ui.adsabs.harvard.edu/abs/2013PASP..125..306F} {125, 306}

\bibitem[\protect\citeauthoryear{{F{\"o}rster Schreiber} et~al.,}{{F{\"o}rster
  Schreiber} et~al.}{2018}]{forster18}
{F{\"o}rster Schreiber} N.~M.,  et~al., 2018, \mn@doi [\apjs]
  {10.3847/1538-4365/aadd49}, \href
  {https://ui.adsabs.harvard.edu/abs/2018ApJS..238...21F} {238, 21}

\bibitem[\protect\citeauthoryear{{Fraternali} \& {Binney}}{{Fraternali} \&
  {Binney}}{2006}]{fraternali06}
{Fraternali} F.,  {Binney} J.~J.,  2006, \mn@doi [\mnras]
  {10.1111/j.1365-2966.2005.09816.x}, \href
  {https://ui.adsabs.harvard.edu/abs/2006MNRAS.366..449F} {366, 449}

\bibitem[\protect\citeauthoryear{{Fraternali}, {van Moorsel}, {Sancisi}  \&
  {Oosterloo}}{{Fraternali} et~al.}{2002}]{fraternali02}
{Fraternali} F.,  {van Moorsel} G.,  {Sancisi} R.,   {Oosterloo} T.,  2002,
  \mn@doi [\aj] {10.1086/340358}, \href
  {https://ui.adsabs.harvard.edu/abs/2002AJ....123.3124F} {123, 3124}

\bibitem[\protect\citeauthoryear{{Fraternali}, {Karim}, {Magnelli},
  {G{\'o}mez-Guijarro}, {Jim{\'e}nez-Andrade}  \& {Posses}}{{Fraternali}
  et~al.}{2021}]{fraternali21}
{Fraternali} F.,  {Karim} A.,  {Magnelli} B.,  {G{\'o}mez-Guijarro} C.,
  {Jim{\'e}nez-Andrade} E.~F.,   {Posses} A.~C.,  2021, \mn@doi [\aap]
  {10.1051/0004-6361/202039807}, \href
  {https://ui.adsabs.harvard.edu/abs/2021A&A...647A.194F} {647, A194}

\bibitem[\protect\citeauthoryear{{Fujimoto} et~al.,}{{Fujimoto}
  et~al.}{2020}]{fujimoto20}
{Fujimoto} S.,  et~al., 2020, \mn@doi [\apj] {10.3847/1538-4357/ab94b3}, \href
  {https://ui.adsabs.harvard.edu/abs/2020ApJ...900....1F} {900, 1}

\bibitem[\protect\citeauthoryear{{Ginsburg} et~al.,}{{Ginsburg}
  et~al.}{2019}]{ginsburg19}
{Ginsburg} A.,  et~al., 2019, \mn@doi [\aj] {10.3847/1538-3881/aafc33}, \href
  {https://ui.adsabs.harvard.edu/abs/2019AJ....157...98G} {157, 98}

\bibitem[\protect\citeauthoryear{{Goodman} \& {Weare}}{{Goodman} \&
  {Weare}}{2010}]{goodman10}
{Goodman} J.,  {Weare} J.,  2010, \mn@doi [Communications in Applied
  Mathematics and Computational Science] {10.2140/camcos.2010.5.65}, \href
  {https://ui.adsabs.harvard.edu/abs/2010CAMCS...5...65G} {5, 65}

\bibitem[\protect\citeauthoryear{{Guilloteau}, {Omont}, {McMahon}, {Cox}  \&
  {Petitjean}}{{Guilloteau} et~al.}{1997}]{guilloteau97}
{Guilloteau} S.,  {Omont} A.,  {McMahon} R.~G.,  {Cox} P.,   {Petitjean} P.,
  1997, \aap, \href {https://ui.adsabs.harvard.edu/abs/1997A&A...328L...1G}
  {328, L1}

\bibitem[\protect\citeauthoryear{{Gullberg} et~al.,}{{Gullberg}
  et~al.}{2015}]{gullberg15}
{Gullberg} B.,  et~al., 2015, \mn@doi [\mnras] {10.1093/mnras/stv372}, \href
  {https://ui.adsabs.harvard.edu/abs/2015MNRAS.449.2883G} {449, 2883}

\bibitem[\protect\citeauthoryear{{Gullberg} et~al.,}{{Gullberg}
  et~al.}{2018}]{gullberg18}
{Gullberg} B.,  et~al., 2018, \mn@doi [\apj] {10.3847/1538-4357/aabe8c}, \href
  {https://ui.adsabs.harvard.edu/abs/2018ApJ...859...12G} {859, 12}

\bibitem[\protect\citeauthoryear{{Gurvich} et~al.,}{{Gurvich}
  et~al.}{2022}]{gurvich22}
{Gurvich} A.~B.,  et~al., 2022, arXiv e-prints, \href
  {https://ui.adsabs.harvard.edu/abs/2022arXiv220304321G} {p. arXiv:2203.04321}

\bibitem[\protect\citeauthoryear{{Hayward} \& {Hopkins}}{{Hayward} \&
  {Hopkins}}{2017}]{hayward17}
{Hayward} C.~C.,  {Hopkins} P.~F.,  2017, \mn@doi [\mnras]
  {10.1093/mnras/stw2888}, \href
  {https://ui.adsabs.harvard.edu/abs/2017MNRAS.465.1682H} {465, 1682}

\bibitem[\protect\citeauthoryear{{Herrera-Camus} et~al.,}{{Herrera-Camus}
  et~al.}{2022}]{herreracamus22}
{Herrera-Camus} R.,  et~al., 2022, \mn@doi [\aap]
  {10.1051/0004-6361/202142562}, \href
  {https://ui.adsabs.harvard.edu/abs/2022A&A...665L...8H} {665, L8}

\bibitem[\protect\citeauthoryear{{Hodge} \& {da Cunha}}{{Hodge} \& {da
  Cunha}}{2020}]{hodge20}
{Hodge} J.~A.,  {da Cunha} E.,  2020, \mn@doi [Royal Society Open Science]
  {10.1098/rsos.200556}, \href
  {https://ui.adsabs.harvard.edu/abs/2020RSOS....700556H} {7, 200556}

\bibitem[\protect\citeauthoryear{{Hodge}, {Carilli}, {Walter}, {de Blok},
  {Riechers}, {Daddi}  \& {Lentati}}{{Hodge} et~al.}{2012}]{hodge12}
{Hodge} J.~A.,  {Carilli} C.~L.,  {Walter} F.,  {de Blok} W.~J.~G.,  {Riechers}
  D.,  {Daddi} E.,   {Lentati} L.,  2012, \mn@doi [\apj]
  {10.1088/0004-637X/760/1/11}, \href
  {https://ui.adsabs.harvard.edu/abs/2012ApJ...760...11H} {760, 11}

\bibitem[\protect\citeauthoryear{{Holwerda}, {Pirzkal}, {de Blok}, {Bouchard},
  {Blyth}  \& {van der Heyden}}{{Holwerda} et~al.}{2011}]{holwerda11}
{Holwerda} B.~W.,  {Pirzkal} N.,  {de Blok} W.~J.~G.,  {Bouchard} A.,  {Blyth}
  S.~L.,   {van der Heyden} K.~J.,  2011, \mn@doi [\mnras]
  {10.1111/j.1365-2966.2011.18942.x}, \href
  {https://ui.adsabs.harvard.edu/abs/2011MNRAS.416.2437H} {416, 2437}

\bibitem[\protect\citeauthoryear{{Iono} et~al.,}{{Iono} et~al.}{2016}]{iono16}
{Iono} D.,  et~al., 2016, \mn@doi [\apjl] {10.3847/2041-8205/829/1/L10}, \href
  {https://ui.adsabs.harvard.edu/abs/2016ApJ...829L..10I} {829, L10}

\bibitem[\protect\citeauthoryear{{Iorio}, {Fraternali}, {Nipoti}, {Di Teodoro},
  {Read}  \& {Battaglia}}{{Iorio} et~al.}{2017}]{iorio17}
{Iorio} G.,  {Fraternali} F.,  {Nipoti} C.,  {Di Teodoro} E.,  {Read} J.~I.,
  {Battaglia} G.,  2017, \mn@doi [\mnras] {10.1093/mnras/stw3285}, \href
  {https://ui.adsabs.harvard.edu/abs/2017MNRAS.466.4159I} {466, 4159}

\bibitem[\protect\citeauthoryear{{Irwin}, {McMahon}  \& {Hazard}}{{Irwin}
  et~al.}{1991}]{irwin91}
{Irwin} M.,  {McMahon} R.~G.,   {Hazard} C.,  1991, in {Crampton} D.,  ed.,
  Astronomical Society of the Pacific Conference Series Vol. 21, The Space
  Distribution of Quasars. pp 117--126

\bibitem[\protect\citeauthoryear{{Jones} et~al.,}{{Jones}
  et~al.}{2016}]{jones16}
{Jones} G.~C.,  et~al., 2016, \mn@doi [\apj] {10.3847/0004-637X/830/2/63},
  \href {https://ui.adsabs.harvard.edu/abs/2016ApJ...830...63J} {830, 63}

\bibitem[\protect\citeauthoryear{{Jones} et~al.,}{{Jones}
  et~al.}{2017}]{jones17}
{Jones} G.~C.,  et~al., 2017, \mn@doi [\apj] {10.3847/1538-4357/aa8df2}, \href
  {https://ui.adsabs.harvard.edu/abs/2017ApJ...850..180J} {850, 180}

\bibitem[\protect\citeauthoryear{{Jones} et~al.,}{{Jones}
  et~al.}{2021}]{jones21}
{Jones} G.~C.,  et~al., 2021, \mn@doi [\mnras] {10.1093/mnras/stab2226}, \href
  {https://ui.adsabs.harvard.edu/abs/2021MNRAS.507.3540J} {507, 3540}

\bibitem[\protect\citeauthoryear{{Kaasinen} et~al.,}{{Kaasinen}
  et~al.}{2020}]{kaasinen20}
{Kaasinen} M.,  et~al., 2020, \mn@doi [\apj] {10.3847/1538-4357/aba438}, \href
  {https://ui.adsabs.harvard.edu/abs/2020ApJ...899...37K} {899, 37}

\bibitem[\protect\citeauthoryear{{Kohandel}, {Pallottini}, {Ferrara},
  {Zanella}, {Behrens}, {Carniani}, {Gallerani}  \& {Vallini}}{{Kohandel}
  et~al.}{2019}]{kohandel19}
{Kohandel} M.,  {Pallottini} A.,  {Ferrara} A.,  {Zanella} A.,  {Behrens} C.,
  {Carniani} S.,  {Gallerani} S.,   {Vallini} L.,  2019, \mn@doi [\mnras]
  {10.1093/mnras/stz1486}, \href
  {https://ui.adsabs.harvard.edu/abs/2019MNRAS.487.3007K} {487, 3007}

\bibitem[\protect\citeauthoryear{{Kohandel}, {Pallottini}, {Ferrara},
  {Carniani}, {Gallerani}, {Vallini}, {Zanella}  \& {Behrens}}{{Kohandel}
  et~al.}{2020}]{kohandel20}
{Kohandel} M.,  {Pallottini} A.,  {Ferrara} A.,  {Carniani} S.,  {Gallerani}
  S.,  {Vallini} L.,  {Zanella} A.,   {Behrens} C.,  2020, \mn@doi [\mnras]
  {10.1093/mnras/staa2792}, \href
  {https://ui.adsabs.harvard.edu/abs/2020MNRAS.499.1250K} {499, 1250}

\bibitem[\protect\citeauthoryear{{Krajnovi{\'c}}, {Cappellari}, {de Zeeuw}  \&
  {Copin}}{{Krajnovi{\'c}} et~al.}{2006}]{krajnovic06}
{Krajnovi{\'c}} D.,  {Cappellari} M.,  {de Zeeuw} P.~T.,   {Copin} Y.,  2006,
  \mn@doi [\mnras] {10.1111/j.1365-2966.2005.09902.x}, \href
  {https://ui.adsabs.harvard.edu/abs/2006MNRAS.366..787K} {366, 787}

\bibitem[\protect\citeauthoryear{{Kretschmer}, {Dekel}  \&
  {Teyssier}}{{Kretschmer} et~al.}{2022}]{kretschmer22}
{Kretschmer} M.,  {Dekel} A.,   {Teyssier} R.,  2022, \mn@doi [\mnras]
  {10.1093/mnras/stab3648}, \href
  {https://ui.adsabs.harvard.edu/abs/2022MNRAS.510.3266K} {510, 3266}

\bibitem[\protect\citeauthoryear{{Krumholz}, {Burkhart}, {Forbes}  \&
  {Crocker}}{{Krumholz} et~al.}{2018}]{krumholz18}
{Krumholz} M.~R.,  {Burkhart} B.,  {Forbes} J.~C.,   {Crocker} R.~M.,  2018,
  \mn@doi [\mnras] {10.1093/mnras/sty852}, \href
  {https://ui.adsabs.harvard.edu/abs/2018MNRAS.477.2716K} {477, 2716}

\bibitem[\protect\citeauthoryear{{Lagache}, {Cousin}  \& {Chatzikos}}{{Lagache}
  et~al.}{2018}]{lagache18}
{Lagache} G.,  {Cousin} M.,   {Chatzikos} M.,  2018, \mn@doi [\aap]
  {10.1051/0004-6361/201732019}, \href
  {https://ui.adsabs.harvard.edu/abs/2018A&A...609A.130L} {609, A130}

\bibitem[\protect\citeauthoryear{{Le F{\`e}vre} et~al.,}{{Le F{\`e}vre}
  et~al.}{2020}]{lefevre20}
{Le F{\`e}vre} O.,  et~al., 2020, \mn@doi [\aap] {10.1051/0004-6361/201936965},
  \href {https://ui.adsabs.harvard.edu/abs/2020A&A...643A...1L} {643, A1}

\bibitem[\protect\citeauthoryear{{Lelli}, {Di Teodoro}, {Fraternali}, {Man},
  {Zhang}, {De Breuck}, {Davis}  \& {Maiolino}}{{Lelli} et~al.}{2021}]{lelli21}
{Lelli} F.,  {Di Teodoro} E.~M.,  {Fraternali} F.,  {Man} A. W.~S.,  {Zhang}
  Z.-Y.,  {De Breuck} C.,  {Davis} T.~A.,   {Maiolino} R.,  2021, \mn@doi
  [Science] {10.1126/science.abc1893}, \href
  {https://ui.adsabs.harvard.edu/abs/2021Sci...371..713L} {371, 713}

\bibitem[\protect\citeauthoryear{{Li} \& {Zhao}}{{Li} \& {Zhao}}{2017}]{li17}
{Li} C.,  {Zhao} G.,  2017, \mn@doi [\apj] {10.3847/1538-4357/aa93f4}, \href
  {https://ui.adsabs.harvard.edu/abs/2017ApJ...850...25L} {850, 25}

\bibitem[\protect\citeauthoryear{{Lu} et~al.,}{{Lu} et~al.}{2017}]{lu17}
{Lu} N.,  et~al., 2017, \mn@doi [\apjs] {10.3847/1538-4365/aa6476}, \href
  {https://ui.adsabs.harvard.edu/abs/2017ApJS..230....1L} {230, 1}

\bibitem[\protect\citeauthoryear{{Lu} et~al.,}{{Lu} et~al.}{2018}]{lu18}
{Lu} N.,  et~al., 2018, \mn@doi [\apj] {10.3847/1538-4357/aad3c9}, \href
  {https://ui.adsabs.harvard.edu/abs/2018ApJ...864...38L} {864, 38}

\bibitem[\protect\citeauthoryear{{Madau} \& {Dickinson}}{{Madau} \&
  {Dickinson}}{2014}]{madau&dickinson14}
{Madau} P.,  {Dickinson} M.,  2014, \mn@doi [\araa]
  {10.1146/annurev-astro-081811-125615}, \href
  {https://ui.adsabs.harvard.edu/abs/2014ARA&A..52..415M} {52, 415}

\bibitem[\protect\citeauthoryear{{Marasco} et~al.,}{{Marasco}
  et~al.}{2019}]{marasco19}
{Marasco} A.,  et~al., 2019, \mn@doi [\aap] {10.1051/0004-6361/201936338},
  \href {https://ui.adsabs.harvard.edu/abs/2019A&A...631A..50M} {631, A50}

\bibitem[\protect\citeauthoryear{{McMullin}, {Waters}, {Schiebel}, {Young}  \&
  {Golap}}{{McMullin} et~al.}{2007}]{mcmullin07}
{McMullin} J.~P.,  {Waters} B.,  {Schiebel} D.,  {Young} W.,   {Golap} K.,
  2007, in {Shaw} R.~A.,  {Hill} F.,   {Bell} D.~J.,  eds,  Astronomical
  Society of the Pacific Conference Series Vol. 376, Astronomical Data Analysis
  Software and Systems XVI. p.~127

\bibitem[\protect\citeauthoryear{{Narayan} \& {Jog}}{{Narayan} \&
  {Jog}}{2002}]{narayan02}
{Narayan} C.~A.,  {Jog} C.~J.,  2002, \mn@doi [\aap]
  {10.1051/0004-6361:20021128}, \href
  {https://ui.adsabs.harvard.edu/abs/2002A&A...394...89N} {394, 89}

\bibitem[\protect\citeauthoryear{{Neeleman}, {Kanekar}, {Prochaska},
  {Rafelski}, {Carilli}  \& {Wolfe}}{{Neeleman} et~al.}{2017}]{neeleman17}
{Neeleman} M.,  {Kanekar} N.,  {Prochaska} J.~X.,  {Rafelski} M.,  {Carilli}
  C.~L.,   {Wolfe} A.~M.,  2017, \mn@doi [Science] {10.1126/science.aal1737},
  \href {https://ui.adsabs.harvard.edu/abs/2017Sci...355.1285N} {355, 1285}

\bibitem[\protect\citeauthoryear{{Neeleman}, {Prochaska}, {Kanekar}  \&
  {Rafelski}}{{Neeleman} et~al.}{2020a}]{neeleman20qubefit}
{Neeleman} M.,  {Prochaska} J.~X.,  {Kanekar} N.,   {Rafelski} M.,  2020a,
  {qubefit: MCMC kinematic modeling}, Astrophysics Source Code Library, record
  ascl:2005.013 (\mn@eprint {ascl} {2005.013})

\bibitem[\protect\citeauthoryear{{Neeleman}, {Prochaska}, {Kanekar}  \&
  {Rafelski}}{{Neeleman} et~al.}{2020b}]{neeleman20}
{Neeleman} M.,  {Prochaska} J.~X.,  {Kanekar} N.,   {Rafelski} M.,  2020b,
  \mn@doi [\nat] {10.1038/s41586-020-2276-y}, \href
  {https://ui.adsabs.harvard.edu/abs/2020Natur.581..269N} {581, 269}

\bibitem[\protect\citeauthoryear{{Neeleman} et~al.,}{{Neeleman}
  et~al.}{2021}]{neeleman21}
{Neeleman} M.,  et~al., 2021, \mn@doi [\apj] {10.3847/1538-4357/abe70f}, \href
  {https://ui.adsabs.harvard.edu/abs/2021ApJ...911..141N} {911, 141}

\bibitem[\protect\citeauthoryear{{Nelson} et~al.,}{{Nelson}
  et~al.}{2019}]{nelson19}
{Nelson} D.,  et~al., 2019, \mn@doi [\mnras] {10.1093/mnras/stz2306}, \href
  {https://ui.adsabs.harvard.edu/abs/2019MNRAS.490.3234N} {490, 3234}

\bibitem[\protect\citeauthoryear{{Noordermeer}, {van der Hulst}, {Sancisi},
  {Swaters}  \& {van Albada}}{{Noordermeer} et~al.}{2005}]{noordermeer05}
{Noordermeer} E.,  {van der Hulst} J.~M.,  {Sancisi} R.,  {Swaters} R.~A.,
  {van Albada} T.~S.,  2005, \mn@doi [\aap] {10.1051/0004-6361:20053172}, \href
  {https://ui.adsabs.harvard.edu/abs/2005A&A...442..137N} {442, 137}

\bibitem[\protect\citeauthoryear{{Oteo} et~al.,}{{Oteo} et~al.}{2016}]{oteo16}
{Oteo} I.,  et~al., 2016, \mn@doi [\apj] {10.3847/0004-637X/827/1/34}, \href
  {https://ui.adsabs.harvard.edu/abs/2016ApJ...827...34O} {827, 34}

\bibitem[\protect\citeauthoryear{{Pallottini} et~al.,}{{Pallottini}
  et~al.}{2022}]{pallottini22}
{Pallottini} A.,  et~al., 2022, \mn@doi [\mnras] {10.1093/mnras/stac1281},
  \href {https://ui.adsabs.harvard.edu/abs/2022MNRAS.513.5621P} {513, 5621}

\bibitem[\protect\citeauthoryear{Pedregosa et~al.,}{Pedregosa
  et~al.}{2011}]{scikit-learn}
Pedregosa F.,  et~al., 2011, Journal of Machine Learning Research, 12, 2825

\bibitem[\protect\citeauthoryear{{Pensabene}, {Carniani}, {Perna}, {Cresci},
  {Decarli}, {Maiolino}  \& {Marconi}}{{Pensabene} et~al.}{2020}]{pensabene20}
{Pensabene} A.,  {Carniani} S.,  {Perna} M.,  {Cresci} G.,  {Decarli} R.,
  {Maiolino} R.,   {Marconi} A.,  2020, \mn@doi [\aap]
  {10.1051/0004-6361/201936634}, \href
  {https://ui.adsabs.harvard.edu/abs/2020A&A...637A..84P} {637, A84}

\bibitem[\protect\citeauthoryear{{Pillepich} et~al.,}{{Pillepich}
  et~al.}{2019}]{pillepich19}
{Pillepich} A.,  et~al., 2019, \mn@doi [\mnras] {10.1093/mnras/stz2338}, \href
  {https://ui.adsabs.harvard.edu/abs/2019MNRAS.490.3196P} {490, 3196}

\bibitem[\protect\citeauthoryear{{Planck Collaboration} et~al.,}{{Planck
  Collaboration} et~al.}{2020}]{planck20}
{Planck Collaboration} et~al., 2020, \mn@doi [\aap]
  {10.1051/0004-6361/201833880}, \href
  {https://ui.adsabs.harvard.edu/abs/2020A&A...641A...1P} {641, A1}

\bibitem[\protect\citeauthoryear{{Ramos Padilla}, {Wang}, {Ploeckinger}, {van
  der Tak}  \& {Trager}}{{Ramos Padilla} et~al.}{2021}]{ramospadilla21}
{Ramos Padilla} A.~F.,  {Wang} L.,  {Ploeckinger} S.,  {van der Tak} F.~F.~S.,
   {Trager} S.~C.,  2021, \mn@doi [\aap] {10.1051/0004-6361/202038207}, \href
  {https://ui.adsabs.harvard.edu/abs/2021A&A...645A.133R} {645, A133}

\bibitem[\protect\citeauthoryear{{Rizzo}, {Vegetti}, {Fraternali}  \& {Di
  Teodoro}}{{Rizzo} et~al.}{2018}]{rizzo18}
{Rizzo} F.,  {Vegetti} S.,  {Fraternali} F.,   {Di Teodoro} E.,  2018, \mn@doi
  [\mnras] {10.1093/mnras/sty2594}, \href
  {https://ui.adsabs.harvard.edu/abs/2018MNRAS.481.5606R} {481, 5606}

\bibitem[\protect\citeauthoryear{{Rizzo}, {Vegetti}, {Powell}, {Fraternali},
  {McKean}, {Stacey}  \& {White}}{{Rizzo} et~al.}{2020}]{rizzo20}
{Rizzo} F.,  {Vegetti} S.,  {Powell} D.,  {Fraternali} F.,  {McKean} J.~P.,
  {Stacey} H.~R.,   {White} S.~D.~M.,  2020, \mn@doi [\nat]
  {10.1038/s41586-020-2572-6}, \href
  {https://ui.adsabs.harvard.edu/abs/2020Natur.584..201R} {584, 201}

\bibitem[\protect\citeauthoryear{{Rizzo}, {Vegetti}, {Fraternali}, {Stacey}  \&
  {Powell}}{{Rizzo} et~al.}{2021}]{rizzo21}
{Rizzo} F.,  {Vegetti} S.,  {Fraternali} F.,  {Stacey} H.~R.,   {Powell} D.,
  2021, \mn@doi [\mnras] {10.1093/mnras/stab2295}, \href
  {https://ui.adsabs.harvard.edu/abs/2021MNRAS.507.3952R} {507, 3952}

\bibitem[\protect\citeauthoryear{{Rizzo}, {Kohandel}, {Pallottini}, {Zanella},
  {Ferrara}, {Vallini}  \& {Toft}}{{Rizzo} et~al.}{2022}]{rizzo22}
{Rizzo} F.,  {Kohandel} M.,  {Pallottini} A.,  {Zanella} A.,  {Ferrara} A.,
  {Vallini} L.,   {Toft} S.,  2022, \mn@doi [\aap]
  {10.1051/0004-6361/202243582}, \href
  {https://ui.adsabs.harvard.edu/abs/2022A&A...667A...5R} {667, A5}

\bibitem[\protect\citeauthoryear{{Robertson} et~al.,}{{Robertson}
  et~al.}{2023}]{robertson23}
{Robertson} B.~E.,  et~al., 2023, \mn@doi [\apjl] {10.3847/2041-8213/aca086},
  \href {https://ui.adsabs.harvard.edu/abs/2023ApJ...942L..42R} {942, L42}

\bibitem[\protect\citeauthoryear{{Romanowsky} \& {Fall}}{{Romanowsky} \&
  {Fall}}{2012}]{romanowsky12}
{Romanowsky} A.~J.,  {Fall} S.~M.,  2012, \mn@doi [\apjs]
  {10.1088/0067-0049/203/2/17}, \href
  {https://ui.adsabs.harvard.edu/abs/2012ApJS..203...17R} {203, 17}

\bibitem[\protect\citeauthoryear{{Shao} et~al.,}{{Shao} et~al.}{2022}]{shao22}
{Shao} Y.,  et~al., 2022, \mn@doi [\aap] {10.1051/0004-6361/202244610}, \href
  {https://ui.adsabs.harvard.edu/abs/2022A&A...668A.121S} {668, A121}

\bibitem[\protect\citeauthoryear{{Sharda} et~al.,}{{Sharda}
  et~al.}{2019}]{sharda19}
{Sharda} P.,  et~al., 2019, \mn@doi [\mnras] {10.1093/mnras/stz1543}, \href
  {https://ui.adsabs.harvard.edu/abs/2019MNRAS.487.4305S} {487, 4305}

\bibitem[\protect\citeauthoryear{{Simons} et~al.,}{{Simons}
  et~al.}{2017}]{simons17}
{Simons} R.~C.,  et~al., 2017, \mn@doi [\apj] {10.3847/1538-4357/aa740c}, \href
  {https://ui.adsabs.harvard.edu/abs/2017ApJ...843...46S} {843, 46}

\bibitem[\protect\citeauthoryear{{Smit} et~al.,}{{Smit} et~al.}{2018}]{smit18}
{Smit} R.,  et~al., 2018, \mn@doi [\nat] {10.1038/nature24631}, \href
  {https://ui.adsabs.harvard.edu/abs/2018Natur.553..178S} {553, 178}

\bibitem[\protect\citeauthoryear{{Somerville} \& {Dav{\'e}}}{{Somerville} \&
  {Dav{\'e}}}{2015}]{somerville15}
{Somerville} R.~S.,  {Dav{\'e}} R.,  2015, \mn@doi [\araa]
  {10.1146/annurev-astro-082812-140951}, \href
  {https://ui.adsabs.harvard.edu/abs/2015ARA&A..53...51S} {53, 51}

\bibitem[\protect\citeauthoryear{{Swaters}, {van Albada}, {van der Hulst}  \&
  {Sancisi}}{{Swaters} et~al.}{2002}]{swaters02}
{Swaters} R.~A.,  {van Albada} T.~S.,  {van der Hulst} J.~M.,   {Sancisi} R.,
  2002, \mn@doi [\aap] {10.1051/0004-6361:20011755}, \href
  {https://ui.adsabs.harvard.edu/abs/2002A&A...390..829S} {390, 829}

\bibitem[\protect\citeauthoryear{{Swinbank} et~al.,}{{Swinbank}
  et~al.}{2017}]{swinbank17}
{Swinbank} A.~M.,  et~al., 2017, \mn@doi [\mnras] {10.1093/mnras/stx201}, \href
  {https://ui.adsabs.harvard.edu/abs/2017MNRAS.467.3140S} {467, 3140}

\bibitem[\protect\citeauthoryear{{Tadaki} et~al.,}{{Tadaki}
  et~al.}{2018}]{tadaki18}
{Tadaki} K.,  et~al., 2018, \mn@doi [\nat] {10.1038/s41586-018-0443-1}, \href
  {https://ui.adsabs.harvard.edu/abs/2018Natur.560..613T} {560, 613}

\bibitem[\protect\citeauthoryear{{Tadaki} et~al.,}{{Tadaki}
  et~al.}{2019}]{tadaki19}
{Tadaki} K.-i.,  et~al., 2019, \mn@doi [\apj] {10.3847/1538-4357/ab1415}, \href
  {https://ui.adsabs.harvard.edu/abs/2019ApJ...876....1T} {876, 1}

\bibitem[\protect\citeauthoryear{{Tadaki} et~al.,}{{Tadaki}
  et~al.}{2020}]{tadaki20}
{Tadaki} K.-i.,  et~al., 2020, \mn@doi [\apj] {10.3847/1538-4357/ab64f4}, \href
  {https://ui.adsabs.harvard.edu/abs/2020ApJ...889..141T} {889, 141}

\bibitem[\protect\citeauthoryear{{Tamburro}, {Rix}, {Leroy}, {Mac Low},
  {Walter}, {Kennicutt}, {Brinks}  \& {de Blok}}{{Tamburro}
  et~al.}{2009}]{tamburro09}
{Tamburro} D.,  {Rix} H.~W.,  {Leroy} A.~K.,  {Mac Low} M.~M.,  {Walter} F.,
  {Kennicutt} R.~C.,  {Brinks} E.,   {de Blok} W.~J.~G.,  2009, \mn@doi [\aj]
  {10.1088/0004-6256/137/5/4424}, \href
  {https://ui.adsabs.harvard.edu/abs/2009AJ....137.4424T} {137, 4424}

\bibitem[\protect\citeauthoryear{{Tamfal}, {Mayer}, {Quinn}, {Babul}, {Madau},
  {Capelo}, {Shen}  \& {Staub}}{{Tamfal} et~al.}{2022}]{tamfal22}
{Tamfal} T.,  {Mayer} L.,  {Quinn} T.~R.,  {Babul} A.,  {Madau} P.,  {Capelo}
  P.~R.,  {Shen} S.,   {Staub} M.,  2022, \mn@doi [\apj]
  {10.3847/1538-4357/ac558e}, \href
  {https://ui.adsabs.harvard.edu/abs/2022ApJ...928..106T} {928, 106}

\bibitem[\protect\citeauthoryear{{Tarantino} et~al.,}{{Tarantino}
  et~al.}{2021}]{tarantino21}
{Tarantino} E.,  et~al., 2021, \mn@doi [\apj] {10.3847/1538-4357/abfcc6}, \href
  {https://ui.adsabs.harvard.edu/abs/2021ApJ...915...92T} {915, 92}

\bibitem[\protect\citeauthoryear{{Toft} et~al.,}{{Toft} et~al.}{2014}]{toft14}
{Toft} S.,  et~al., 2014, \mn@doi [\apj] {10.1088/0004-637X/782/2/68}, \href
  {https://ui.adsabs.harvard.edu/abs/2014ApJ...782...68T} {782, 68}

\bibitem[\protect\citeauthoryear{{Tokuoka} et~al.,}{{Tokuoka}
  et~al.}{2022}]{tokuoka22}
{Tokuoka} T.,  et~al., 2022, \mn@doi [\apjl] {10.3847/2041-8213/ac7447}, \href
  {https://ui.adsabs.harvard.edu/abs/2022ApJ...933L..19T} {933, L19}

\bibitem[\protect\citeauthoryear{{Toomre}}{{Toomre}}{1964}]{toomre64}
{Toomre} A.,  1964, \mn@doi [\apj] {10.1086/147861}, \href
  {https://ui.adsabs.harvard.edu/abs/1964ApJ...139.1217T} {139, 1217}

\bibitem[\protect\citeauthoryear{{Tsukui} \& {Iguchi}}{{Tsukui} \&
  {Iguchi}}{2021}]{tsukui21}
{Tsukui} T.,  {Iguchi} S.,  2021, \mn@doi [Science] {10.1126/science.abe9680},
  372, 1201

\bibitem[\protect\citeauthoryear{{Turner} et~al.,}{{Turner}
  et~al.}{2017}]{turner17}
{Turner} O.~J.,  et~al., 2017, \mn@doi [\mnras] {10.1093/mnras/stx1366}, \href
  {https://ui.adsabs.harvard.edu/abs/2017MNRAS.471.1280T} {471, 1280}

\bibitem[\protect\citeauthoryear{{{\"U}bler} et~al.,}{{{\"U}bler}
  et~al.}{2019}]{uebler19}
{{\"U}bler} H.,  et~al., 2019, \mn@doi [\apj] {10.3847/1538-4357/ab27cc}, \href
  {https://ui.adsabs.harvard.edu/abs/2019ApJ...880...48U} {880, 48}

\bibitem[\protect\citeauthoryear{{Valentino} et~al.,}{{Valentino}
  et~al.}{2020}]{valentino20}
{Valentino} F.,  et~al., 2020, \mn@doi [\apj] {10.3847/1538-4357/ab64dc}, \href
  {https://ui.adsabs.harvard.edu/abs/2020ApJ...889...93V} {889, 93}

\bibitem[\protect\citeauthoryear{{Verheijen} \& {Sancisi}}{{Verheijen} \&
  {Sancisi}}{2001}]{verheijen01}
{Verheijen} M.~A.~W.,  {Sancisi} R.,  2001, \mn@doi [\aap]
  {10.1051/0004-6361:20010090}, \href
  {https://ui.adsabs.harvard.edu/abs/2001A&A...370..765V} {370, 765}

\bibitem[\protect\citeauthoryear{{Wagg}, {Carilli}, {Wilner}, {Cox}, {De
  Breuck}, {Menten}, {Riechers}  \& {Walter}}{{Wagg} et~al.}{2010}]{wagg10}
{Wagg} J.,  {Carilli} C.~L.,  {Wilner} D.~J.,  {Cox} P.,  {De Breuck} C.,
  {Menten} K.,  {Riechers} D.~A.,   {Walter} F.,  2010, \mn@doi [\aap]
  {10.1051/0004-6361/201015424}, \href
  {https://ui.adsabs.harvard.edu/abs/2010A&A...519L...1W} {519, L1}

\bibitem[\protect\citeauthoryear{{Wagg} et~al.,}{{Wagg} et~al.}{2014}]{wagg14}
{Wagg} J.,  et~al., 2014, \mn@doi [\apj] {10.1088/0004-637X/783/2/71}, \href
  {https://ui.adsabs.harvard.edu/abs/2014ApJ...783...71W} {783, 71}

\bibitem[\protect\citeauthoryear{{Weijmans} et~al.,}{{Weijmans}
  et~al.}{2014}]{weijmans14}
{Weijmans} A.-M.,  et~al., 2014, \mn@doi [\mnras] {10.1093/mnras/stu1603},
  \href {https://ui.adsabs.harvard.edu/abs/2014MNRAS.444.3340W} {444, 3340}

\bibitem[\protect\citeauthoryear{{Wolfire}, {Vallini}  \& {Chevance}}{{Wolfire}
  et~al.}{2022}]{wolfire22}
{Wolfire} M.~G.,  {Vallini} L.,   {Chevance} M.,  2022, \mn@doi [\araa]
  {10.1146/annurev-astro-052920-010254}, \href
  {https://ui.adsabs.harvard.edu/abs/2022ARA&A..60..247W} {60, 247}

\bibitem[\protect\citeauthoryear{{Yue} et~al.,}{{Yue} et~al.}{2021}]{yue21}
{Yue} M.,  et~al., 2021, \mn@doi [\apj] {10.3847/1538-4357/ac0af4}, \href
  {https://ui.adsabs.harvard.edu/abs/2021ApJ...917...99Y} {917, 99}

\bibitem[\protect\citeauthoryear{{Yun} et~al.,}{{Yun} et~al.}{2015}]{yun15}
{Yun} M.~S.,  et~al., 2015, \mn@doi [\mnras] {10.1093/mnras/stv1963}, \href
  {https://ui.adsabs.harvard.edu/abs/2015MNRAS.454.3485Y} {454, 3485}

\bibitem[\protect\citeauthoryear{{Zolotov} et~al.,}{{Zolotov}
  et~al.}{2015}]{zolotov15}
{Zolotov} A.,  et~al., 2015, \mn@doi [\mnras] {10.1093/mnras/stv740}, \href
  {https://ui.adsabs.harvard.edu/abs/2015MNRAS.450.2327Z} {450, 2327}

\bibitem[\protect\citeauthoryear{{de Blok} et~al.,}{{de Blok}
  et~al.}{2016}]{deblok16}
{de Blok} W.~J.~G.,  et~al., 2016, \mn@doi [\aj] {10.3847/0004-6256/152/2/51},
  \href {https://ui.adsabs.harvard.edu/abs/2016AJ....152...51D} {152, 51}

\bibitem[\protect\citeauthoryear{{van der Hulst}, {van Albada}  \&
  {Sancisi}}{{van der Hulst} et~al.}{2001}]{vanderhulst01}
{van der Hulst} J.~M.,  {van Albada} T.~S.,   {Sancisi} R.,  2001, in {Hibbard}
  J.~E.,  {Rupen} M.,   {van Gorkom} J.~H.,  eds,  Astronomical Society of the
  Pacific Conference Series Vol. 240, Gas and Galaxy Evolution. p.~451

\makeatother
\end{thebibliography}



\appendix

\section{Kinematic model parameters}\label{app:kinpar}
In Table~\ref{tab:BBpar}, we present the parameters used in \texttt{$^{\text{3D}}$BAROLO} to obtain the kinematic models shown in Section~\ref{sec:kin}.

We note that the velocity channels of AzTEC1 and BRI1335-0417 can be considered fully independent since the native spectral resolution of these data are much higher than the channel widths chosen for the datacubes. This is not the case for the datasets of J081740 and SGP38326, in which we have to consider a spectral resolution 1.21 times the velocity channel width (see Section 5.5.2 of the ALMA Cycle 9 Technical Handbook). This affects the \texttt{LINEAR} parameter used in \texttt{$^{\text{3D}}$BAROLO} which is FWHM/2.355 for AzTEC1 and BRI1335-0417 and 1.21 $\times$ FWHM/2.355 for J081740 and SGP38326, where FWHM is the channel width as given in Table~\ref{tab:data}.

\begin{table*}
\caption{Summary of the \texttt{$^{\text{3D}}$BAROLO} parameters for the fiducial kinematic models. These parameters are described in Section~\ref{sec:barolo} and in the documentation of \texttt{$^{\text{3D}}$BAROLO} (see Data Availability). The central coordinates given here in pixels match the RA and DEC given in Table~\ref{tab:data}. The radial separation between the tilted rings (\texttt{RADSEP}) is given in arcsec. The \texttt{LINEAR} parameter controls the spectral broadening of the instrument and is related to the spectral resolution of the data, more details for the chosen values are in the text.}\label{tab:BBpar}
\begin{tabular}{lccccc}
\hline \hline
            & AzTEC1 & BRI1335-0417 & J081740    & SGP38326-1 & SGP38326-2  \\ \hline
X0          & 82     & 46           & 51         & 51         & 44          \\
Y0          & 76     & 49           & 52         & 46         & 49          \\
RADSEP      & 0.11   & 0.15         & 0.13       & 0.13       & 0.12        \\
NRADII      & 4      & 5            & 4          & 5          & 3           \\
INC         & 39     & 42           & 43         & 42         & 41          \\
PA          & 289    & 10           & 99         & free       & 119         \\
Z0          & 0      & 0            & 0          & 0          & 0           \\
GROWTHCUT   & 2.5    & 2.5          & 2.5        & 2.0        & 3.0         \\
SNRCUT      & 3.5    & 3.0          & 3.5        & 3.0        & 3.5         \\
LINEAR      & 0.43   & 0.43         & 0.51       & 0.51       & 0.51        \\
SIDE        & both   & approaching  & both       & both       & both        \\  \hline
\end{tabular}
\end{table*}

\section{Pseudo Contour Method}\label{app:ps}
In Figures~\ref{fig:data} and~\ref{fig:kinmod} we masked the emission with a pseudo 4$\sigma$-contour for better visualisation as it ensures that the emission within the mask is likely real. The pseudo contour method explained here is required primarily because the number of collapsed channels in the total intensity map is not constant across the map. 
To determine the pseudo 4$\sigma$-contour ($\sigma_{\textrm{pc}}$), we used \texttt{CANNUBI} that follows a similar procedure to \citet{verheijen01}. The software first obtains a 3D mask using \texttt{$^{\text{3D}}$BAROLO} by smoothing the datacubes and performing a 3D search of the object. All pixels outside of the mask are set to zero and all the channels containing emission were added to build the integrated emission map. The noise per pixel ($\sigma_{\mathrm{tot}}$) in this map is determined as:

\begin{equation}\label{eq:pseudo}
    \sigma_{\textrm{tot}} = \sqrt{N} \sigma_{\mathrm{[CII]}},
\end{equation}

\noindent where $N$ is number of channels inside the mask that contributed to that pixel and $\sigma_{\mathrm{[CII]}}$ is the noise per channel (see Table~\ref{tab:data}). This allows CANNUBI to create a SNR map. Then the pseudo NT$\sigma$-contour is calculated by selecting all the pixels in this map with $\textrm{SNR}_{\textrm{tot}}$ values in the interval:

\begin{equation}\label{eq:NT}
    0.95 \times \textrm{NT} < \textrm{SNR}_{\textrm{tot}} < 1.05 \times \textrm{NT},
\end{equation}

\noindent where NT, "Noise Times", is an arbitrary level in the SNR map. We choose the value of NT = 4 for all the galaxies, which defines a pseudo 4$\sigma$-contour in the SNR map and secures that any emission inside it is likely real and part of the system. Finally, CANNUBI calcutes the average value of the pixels in the total intensity map corresponding to those identified by Equation~\ref{eq:NT}.
This is the pseudo 4$\sigma$ flux value. We use this flux to determine the mask used in Figure~\ref{fig:data}.

It is important to note that \texttt{CANNUBI} derives the desired pseudo $\sigma$-contour level utilising the 0th moment maps produced by \texttt{$^{\text{3D}}$BAROLO}, which may have different units depending on the software version. Here, we report the [CII] intensity in Figure~\ref{fig:data} and the pseudo 4$\sigma$-contours in Section~\ref{sec:presentation} in units of Jy km s$^{-1}$, which is native to the version 1.6.1 of \texttt{$^{\text{3D}}$BAROLO} used in this work.

\section{Channel Maps}\label{app:cm}
\begin{figure*}
    \centering
    \includegraphics[width=0.7\textwidth]{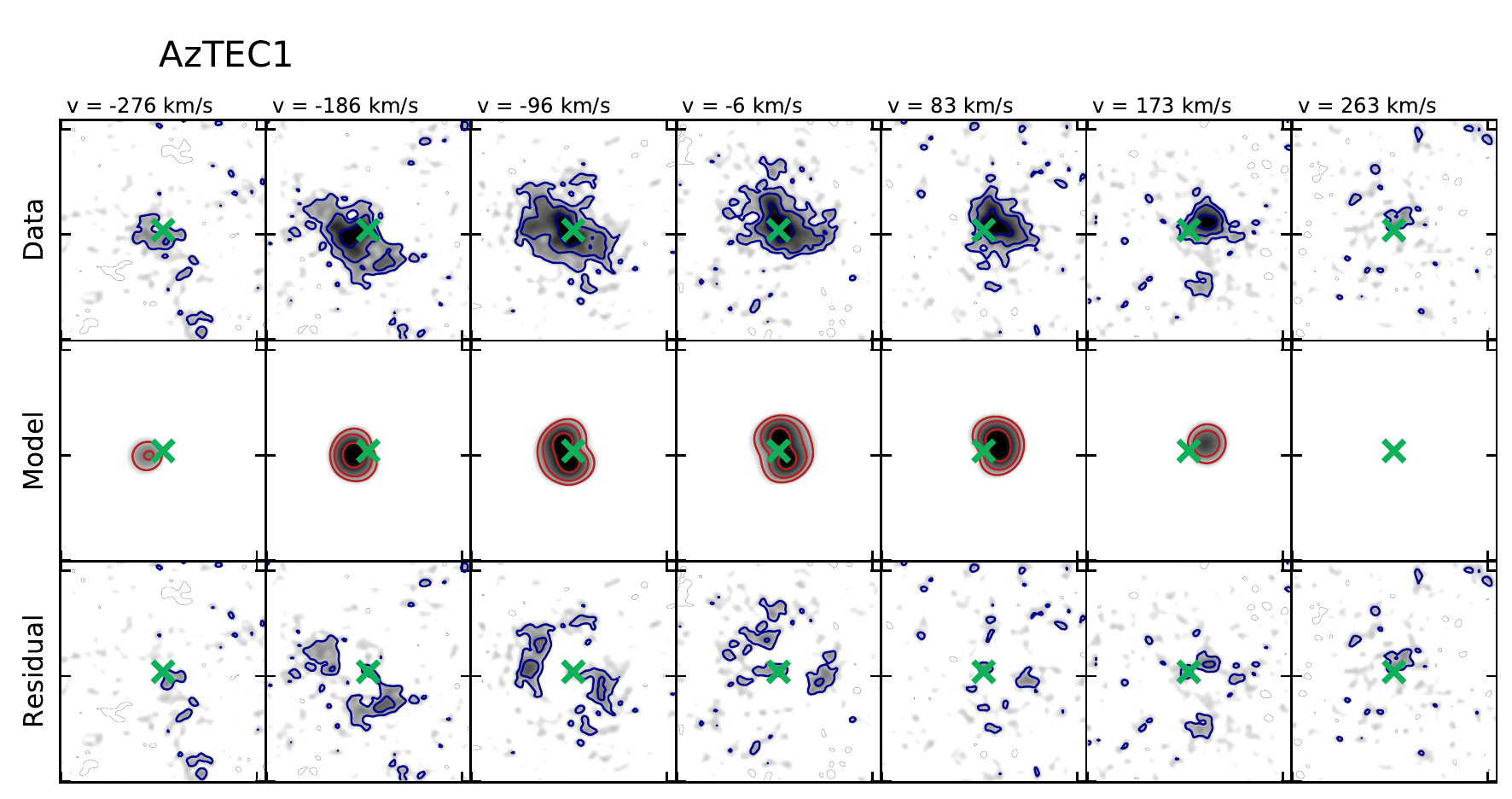}
    \caption{Representative channel maps of AzTEC1. Top panels: [CII] emission data. Middle panels: \texttt{$^{\text{3D}}$BAROLO} rotating disc model. Bottom panels: residuals. The contours follow the emission intensity in $\sigma$ levels ($2\sigma_{\mathrm{[CII]}}, 4\sigma_{\mathrm{[CII]}}, 8\sigma_{\mathrm{[CII]}}$ and $16\sigma_{\mathrm{[CII]}}$ in blue; $-2\sigma_{\mathrm{[CII]}}$ in gray).}
    \label{fig:cmA}
\end{figure*}

\begin{figure*}
    \centering
    \includegraphics[width=0.7\textwidth]{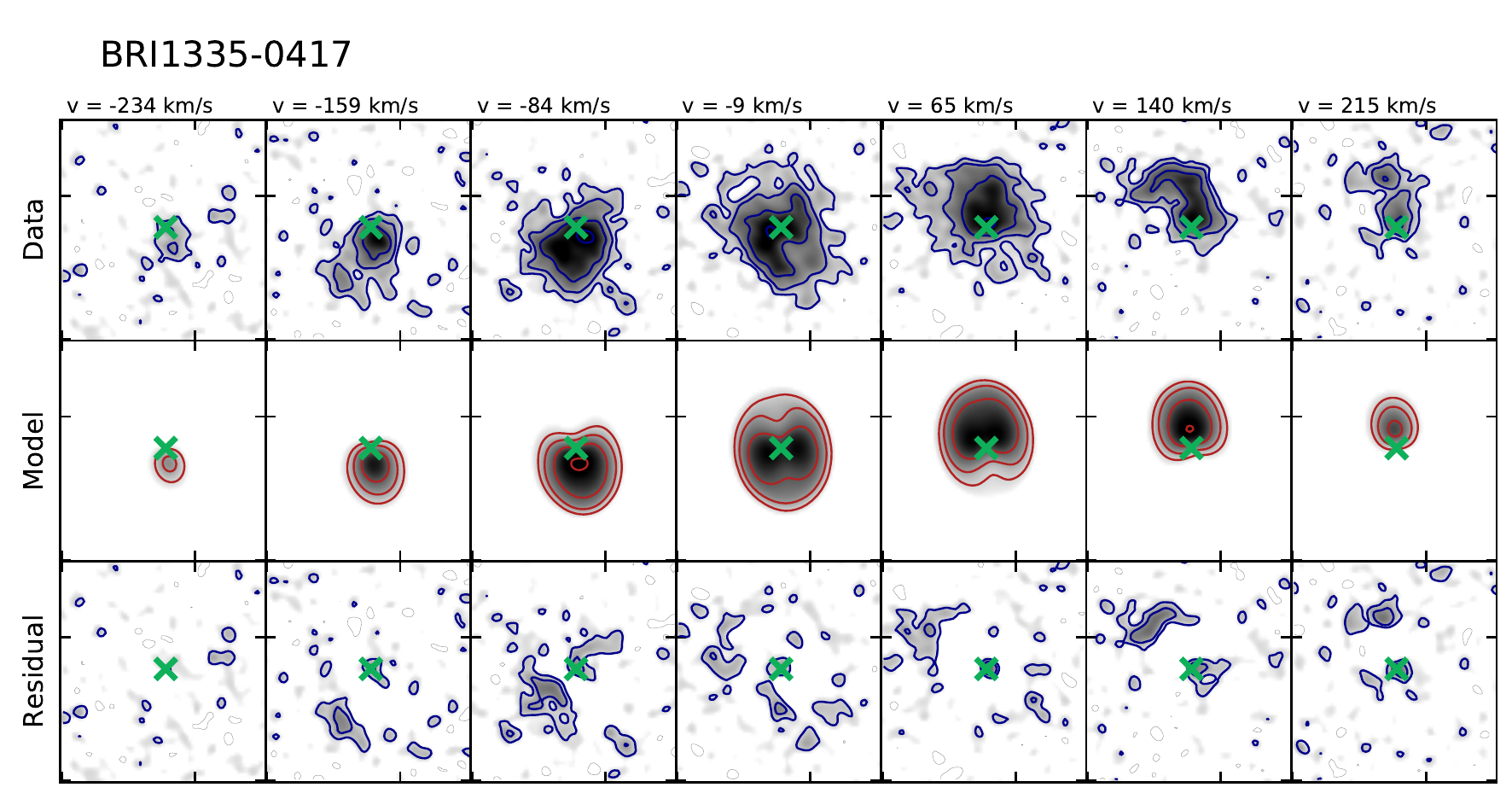}
    \caption{Representative channel maps of BRI1335-0417. The panels are the same as in Figure~\ref{fig:cmA}.}
    \label{fig:cmB}
\end{figure*}

\begin{figure*}
    \centering
    \includegraphics[width=0.7\textwidth]{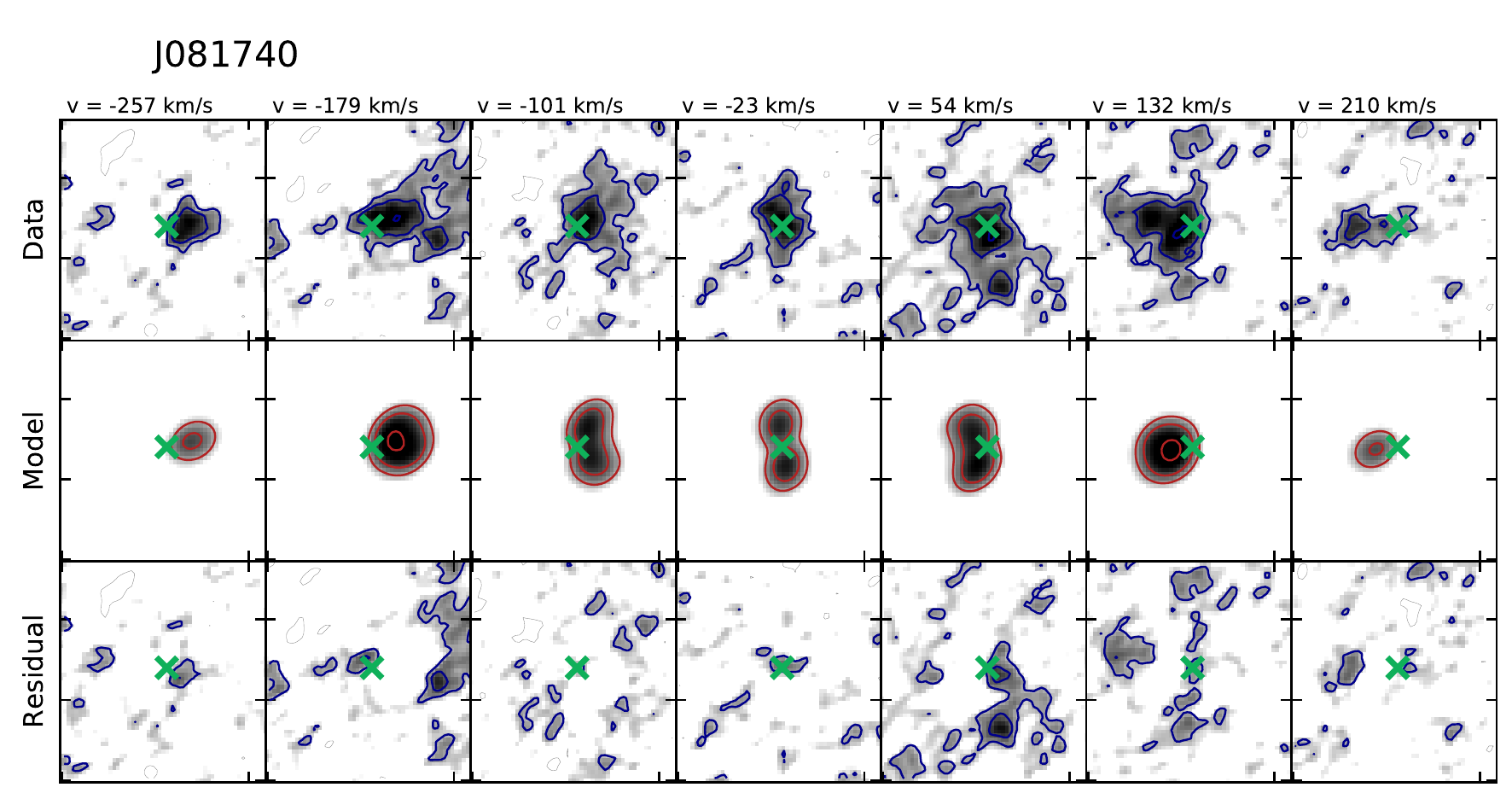}
    \caption{Representative channel maps of J081740. The panels are the same as in Figure~\ref{fig:cmA}.}
    \label{fig:cmJ}
\end{figure*}

\begin{figure*}
    \centering
    \includegraphics[width=0.7\textwidth]{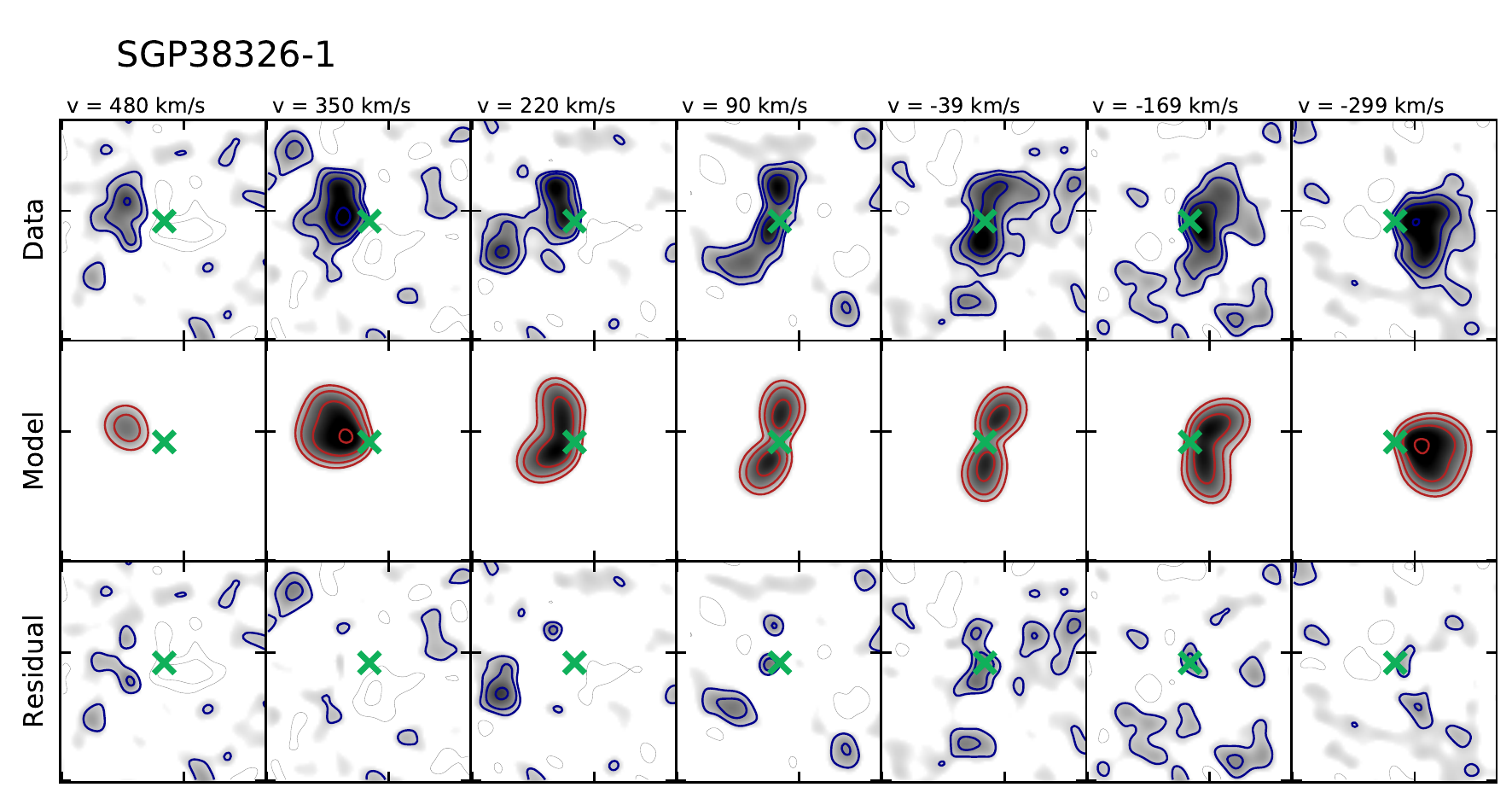}
    \caption{Representative channel maps of SGP38326-1. The panels are the same as in Figure~\ref{fig:cmA}.}
    \label{fig:cmS1}
\end{figure*}

\begin{figure*}
    \centering
    \includegraphics[width=0.7\textwidth]{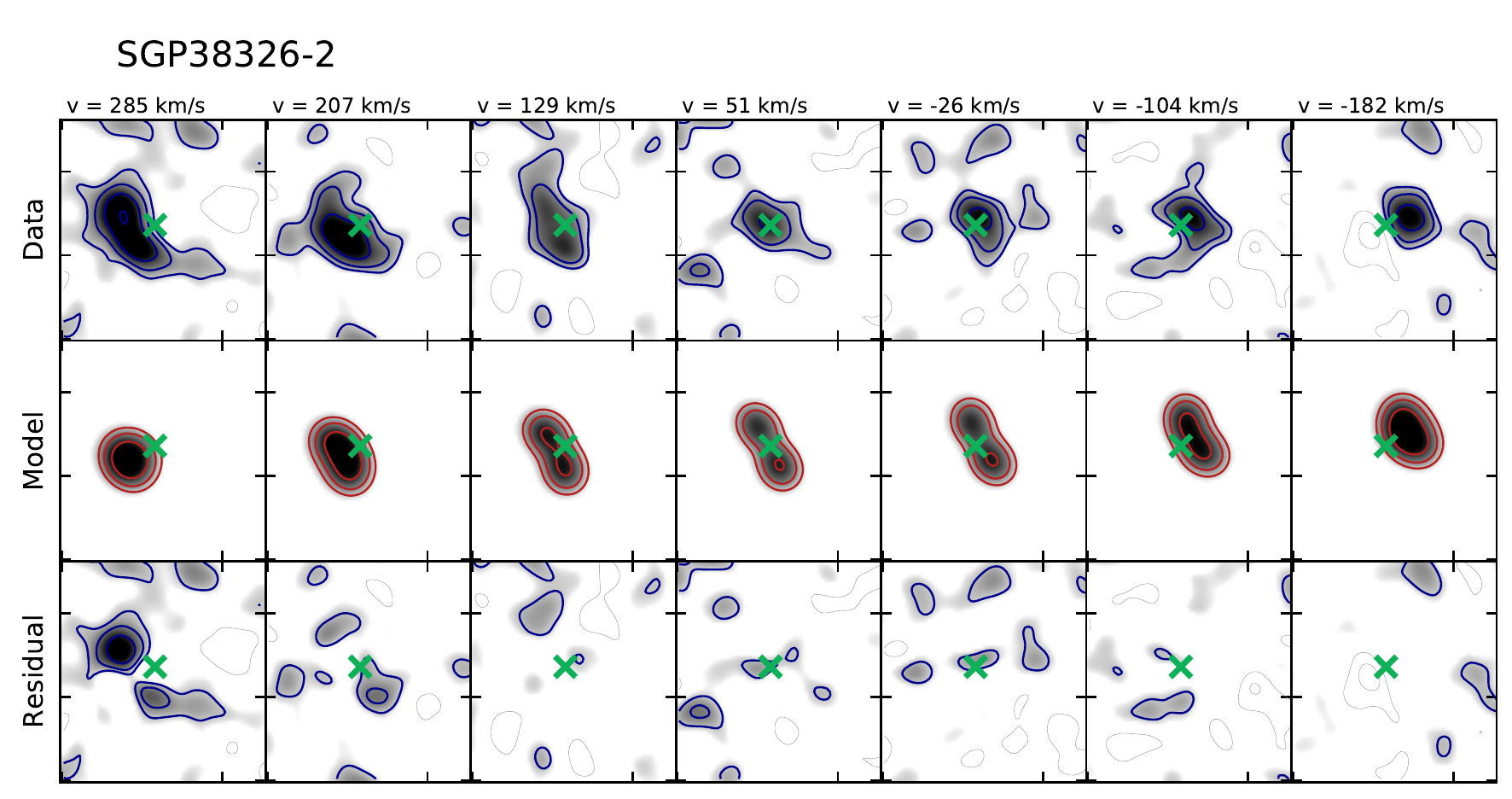}
    \caption{Representative channel maps of SGP38326-2. The panels are the same as in Figure~\ref{fig:cmA}.}
    \label{fig:cmS2}
\end{figure*}

In Figures ~\ref{fig:cmA} - ~\ref{fig:cmS2}, we show representative channel maps of the [CII] emission of each galaxy in our sample alongside the best-fit kinematic model obtained with \texttt{$^{\text{3D}}$BAROLO} that we describe in Section~\ref{sec:kin}. We show three rows of panels, from top to bottom, respectively, the data, the model and the residuals. We show the data in blue contours for the positive emission and gray for the negative emission. The model is shown in red contours. The contours follow the emission intensity according to $2\sigma_{\mathrm{[CII]}}, 4\sigma_{\mathrm{[CII]}}, 8\sigma_{\mathrm{[CII]}}$ and $16\sigma_{\mathrm{[CII]}}$, the negative contours are at $-2\sigma_{\mathrm{[CII]}}$. For the $\sigma_{\mathrm{[CII]}}$ values see Table~\ref{tab:data}.

\section{\texttt{CANNUBI} Corner Plots}\label{app:cannubicorner}

In Figures~\ref{fig:cpA} - ~\ref{fig:cpS2}, we show the posterior distributions obtained by applying \texttt{CANNUBI} on our sample of \textit{z} $\sim 4.5$ galaxies.

\begin{figure*}

\begin{subfigure}{0.5\textwidth}
  \centering
    \includegraphics[width=\textwidth]{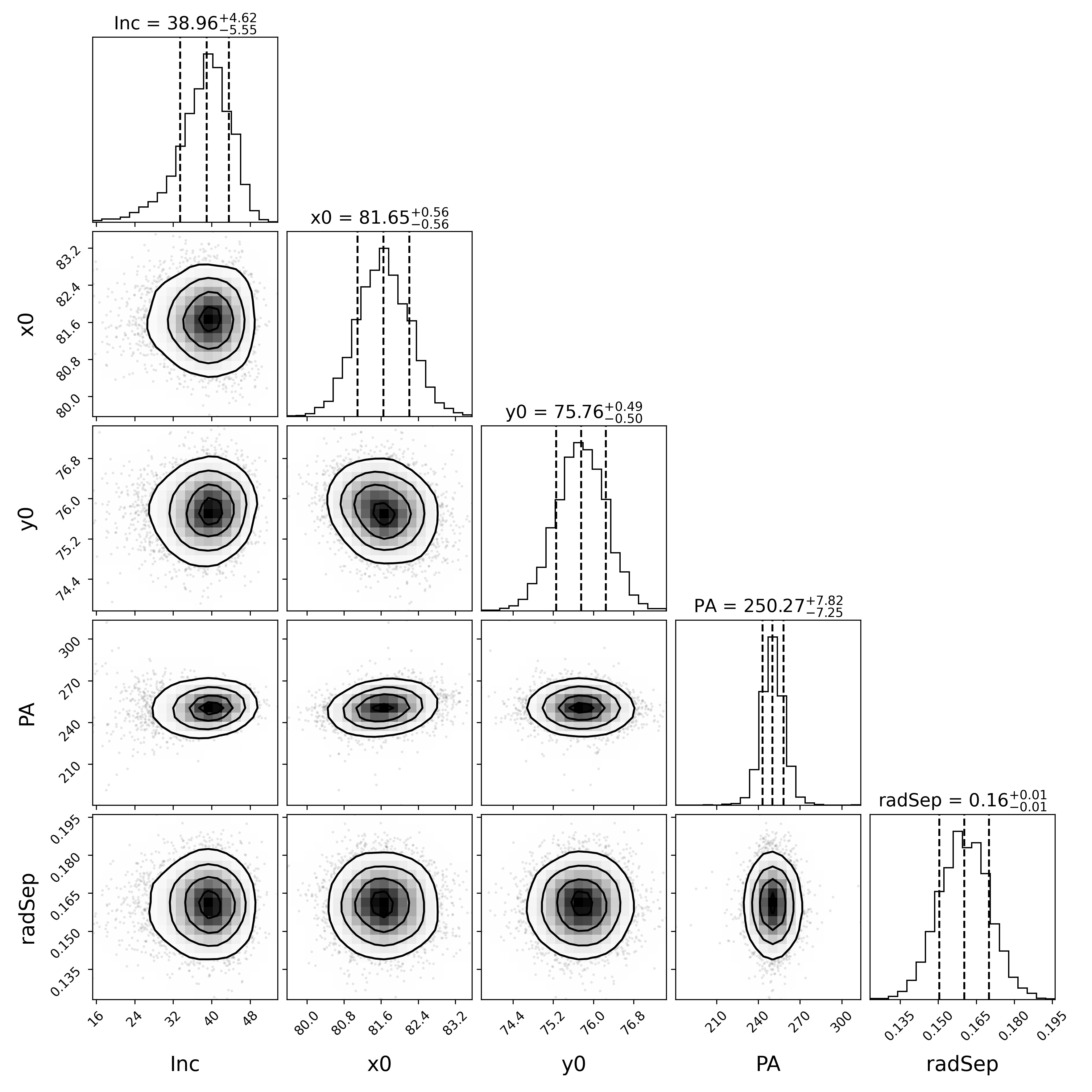}
  \caption{AzTEC1: 4 tilted rings.}
  \label{fig:cpA}
\end{subfigure}\hfill
\begin{subfigure}{0.5\textwidth}
  \centering
    \includegraphics[width=\textwidth]{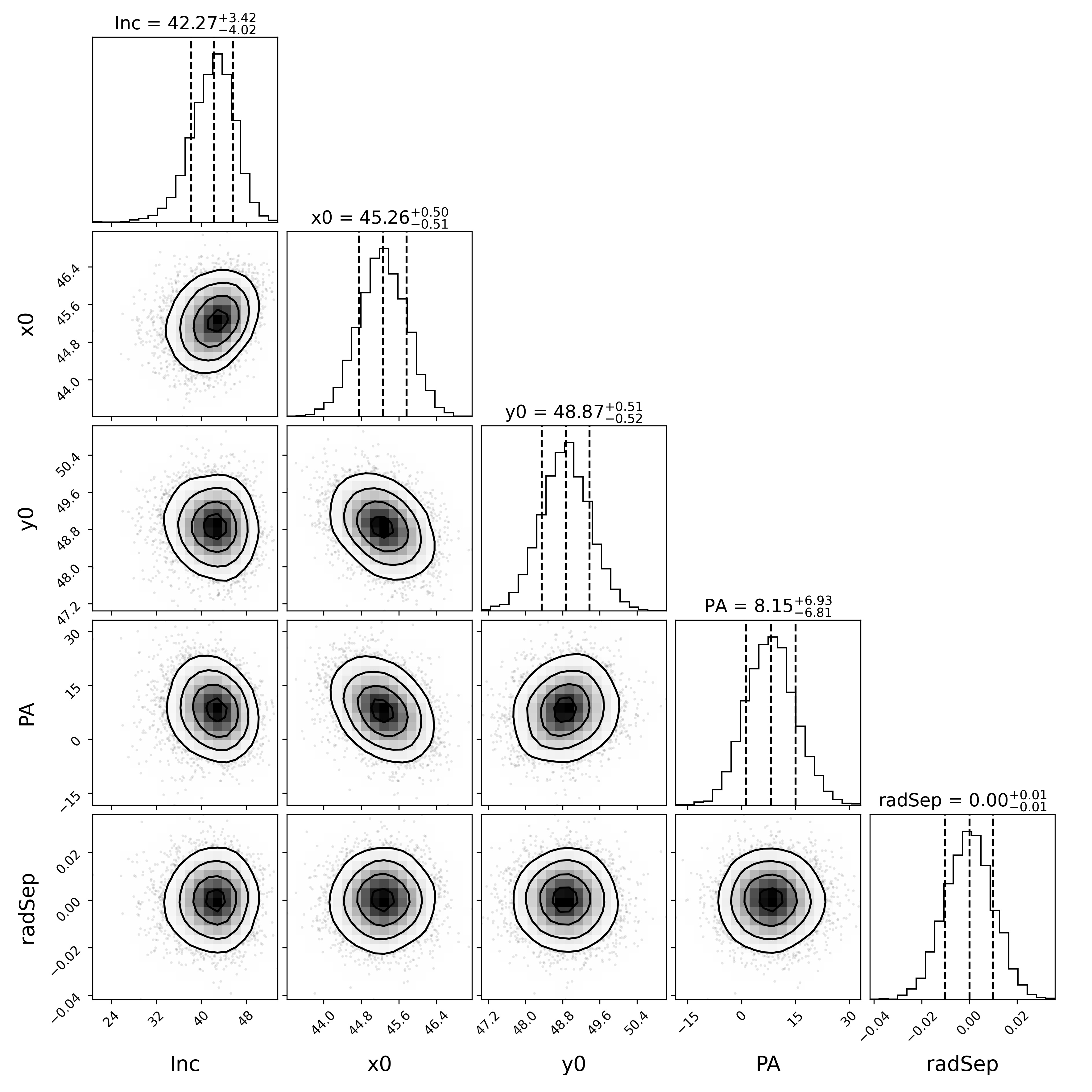}
  \caption{BRI1335-0417: 5 tilted rings.}
  \label{fig:cpB}
\end{subfigure}

\begin{subfigure}{0.5\textwidth}
  \centering
    \includegraphics[width=\textwidth]{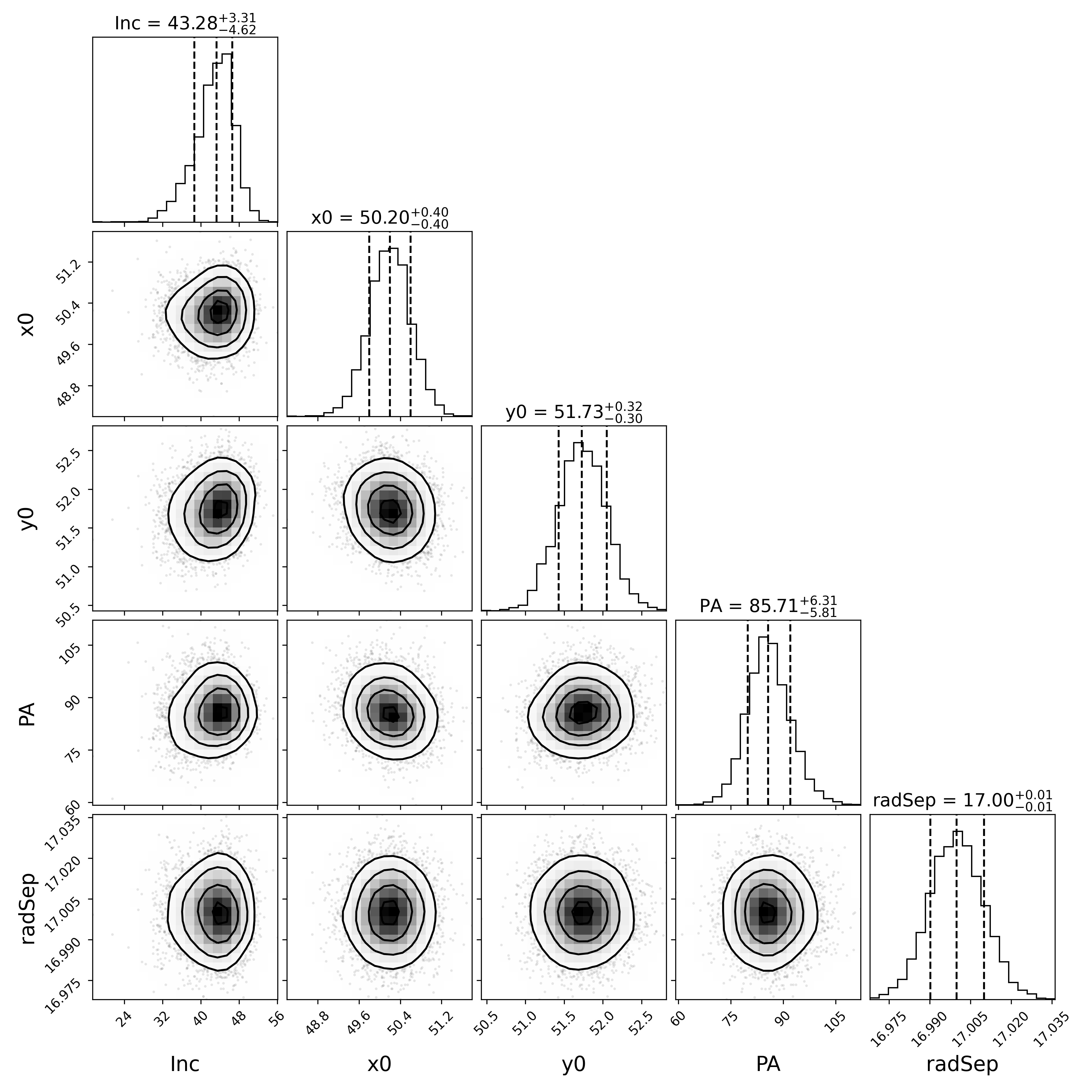}
  \caption{J081740: 4 tilted rings.}
  \label{fig:cpJ}
\end{subfigure}\hfill
\begin{subfigure}{0.5\textwidth}
  \centering
    \includegraphics[width=\textwidth]{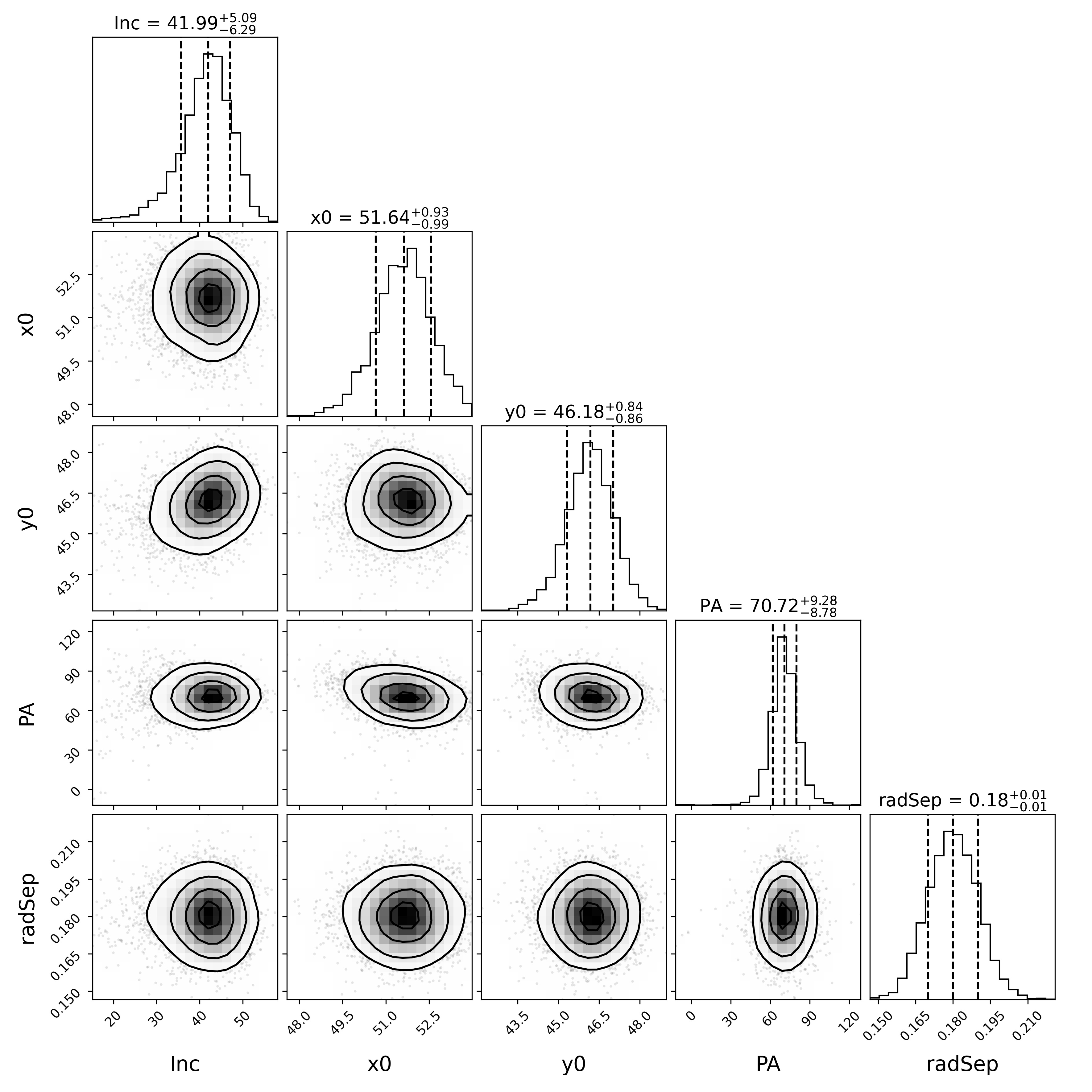}
  \caption{SGP38326-1: 4 tilted rings.}
  \label{fig:cpS1}
\end{subfigure}
\end{figure*}

\begin{figure*}\ContinuedFloat
\centering
\begin{subfigure}{0.5\textwidth}
  \includegraphics[width=\textwidth]{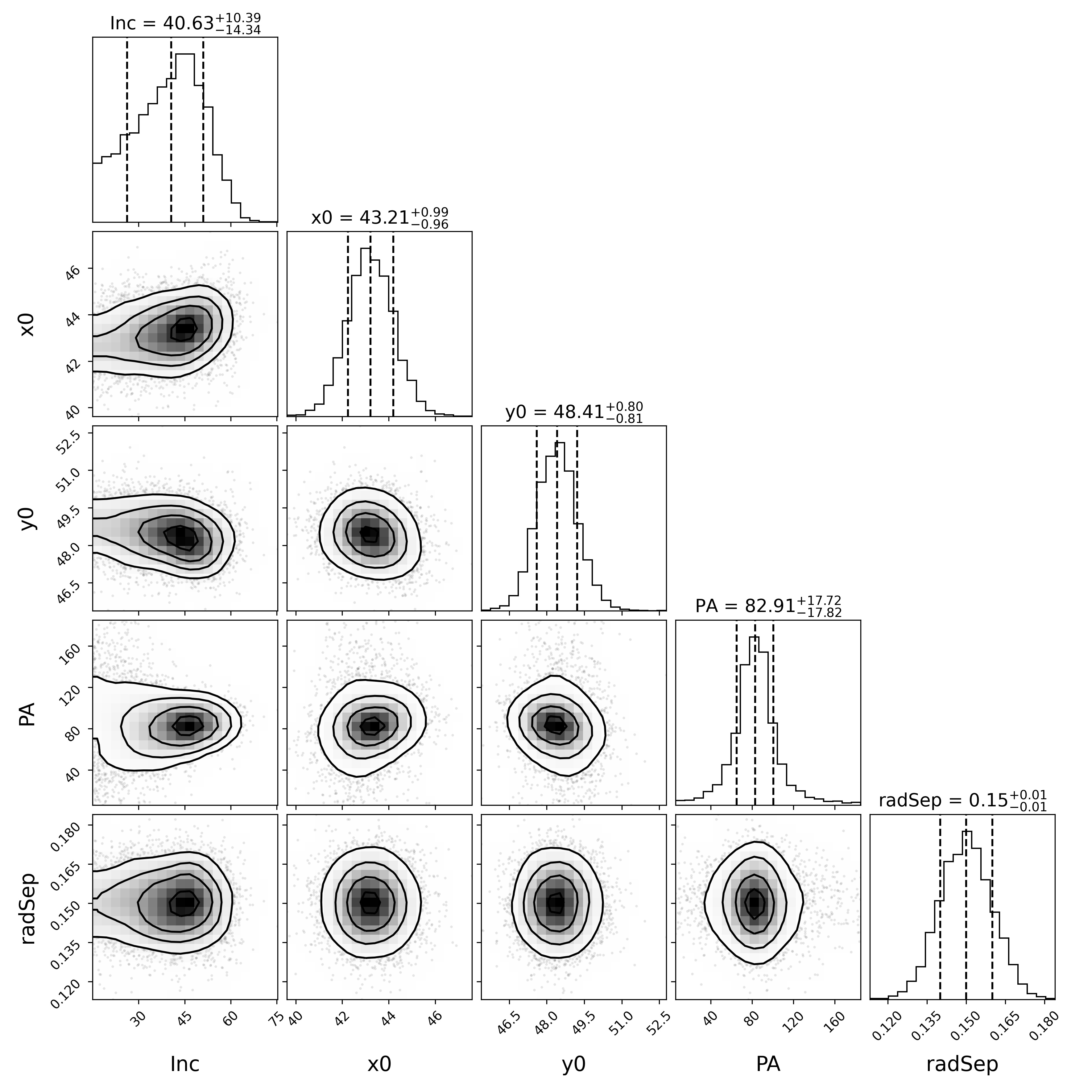}
  \caption{SGP38326-2: 3 tilted rings.}
  \label{fig:cpS2}
\end{subfigure}

\caption{Posterior distributions of \texttt{CANNUBI} on AzTEC1. We display the inclination in degrees, galactic centre in pixels (x0, y0), morphological PA in degrees and radial separation between the tilted rings in arcsec. The best values shown are the median followed by the upper and lower errors defined by the 16th and 84th percentiles (also shown as dashed lines).}
\label{fig:cP}
\end{figure*}


\bsp	
\label{lastpage}

\end{document}